\documentstyle[12pt]{l-aa}
\input{psfig}

\def\Real{{\rm I\mathchoice{\kern-0.70mm}{\kern-0.70mm}{\kern-0.65mm}%
  {\kern-0.50mm}R}}

\font \bolditalics = cmmib10
\def\bx#1{\leavevmode\thinspace\hbox{\vrule\vtop{\vbox{\hrule\kern1pt
        \hbox{\vphantom{\tt/}\thinspace{\bf#1}\thinspace}}
      \kern1pt\hrule}\vrule}\thinspace}

\def \vc #1{{\textfont1=\bolditalics \hbox{$\bf#1$}}}

\def\pg{{\bf p}}

\def\eg{{\bf e}}

\def\thetag{{\vc \theta}}

\def\gammag{{\vc \gamma}}

\def\be{\begin{equation}}
\def\ee{\end{equation}}
\def\ba{\begin{eqnarray}}
\def\ea{\end{eqnarray}}

\begin{document}


   \thesaurus{02         
              (12.03.4;
               12.04.1;
               12.07.1;
               12.12.1)}


   \title{Detection of correlated galaxy ellipticities from CFHT data: first
evidence for gravitational lensing
by large-scale structures \thanks{Based on observations obtained at the
Canada-France-Hawaii Telescope (CFHT) which is operated by the National
Research Council of Canada (NRCC), the Institut des Sciences de l'Univers
(INSU) of the Centre National de la Recherche Scientifique (CNRS) and
the University of Hawaii (UH)}}
   \author{L. Van Waerbeke$^{1}$, Y. Mellier$^{2,3}$, T. Erben$^4$, J.C.
Cuillandre$^5$, F. Bernardeau$^6$,
R. Maoli$^{2,3}$, E. Bertin$^{2,3}$, H.J. Mc Cracken$^7$, O. Le F\`evre$^7$, B.
Fort$^2$,
M. Dantel-Fort$^3$, B. Jain$^{8}$, P. Schneider$^4$}
   \offprints{waerbeke@cita.utoronto.ca}

  \institute{$^1$ Canadian Institut for Theoretical Astrophysics, 60 St 
Georges
Str., Toronto, M5S 3H8 Ontario, Canada.\\
   $^2$ Institut d'Astrophysique de Paris. 98 bis, boulevard
Arago. 75014 Paris, France. \\
   $^3$ Observatoire de Paris. DEMIRM. 61, avenue de
l'Observatoire.  75014 Paris, France.\\
   $^4$ Max Planck Institut fur Astrophysiks, Karl-Schwarzschild-Str. 1,
Postfach 1523,
D-85740 Garching, Germany. \\
   $^5$ Canada-France-Hawaii-Telescope, PO Box 1597, Kamuela, Hawaii 96743,
USA\\
   $^6$ Service de Physique Th\'eorique. C.E. de Saclay. 91191 Gif sur 
Yvette
Cedex, France.\\
   $^7$ Laboratoire d'Astronomie Spatiale, 13376 Marseille Cedex 12, 
France\\
   $^8$ Dept. of Physics, Johns Hopkins University, Baltimore, MD 21218, 
USA\\
}

   \markboth{First detection of cosmic shear}{}


\maketitle
   \markboth{First detection of cosmic shear}{}

\begin{abstract}
We report the detection of a significant (5.5 $\sigma$) excess of 
correlations between galaxy
ellipticities at scales ranging from $0.5$ to $3.5$ arc-minutes.
This detection of a gravitational lensing signal by large-scale
structure was made
using a composite high quality imaging survey of $6300$ arcmin$^2$
obtained at the
Canada France Hawaii Telescope (CFHT) with the
UH8K and CFH12K panoramic CCD cameras.
The amplitude of the excess correlation is $2.2\pm 0.2~\%$ at $1~{\rm arcmin}$
scale, in agreement with theoretical predictions of
the lensing effect induced by large-scale structure.
We provide a quantitative analysis of systematics which could contribute to 
the
signal and show that the net effect is small and can be corrected for.
In particular, we show that the spurious excess of
correlations caused by the residual of the
anisotropic Point Spread Function (PSF) correction
is well below the measured signal.
We show that the measured ellipticity correlations behave as
expected for a gravitational shear signal.
The relatively small size of our survey precludes tight constraints on
cosmological models. However the data are in favor of cluster 
normalized
cosmological models, and marginally reject Cold Dark Matter models with
($\Omega=0.3$, $\sigma_8<0.6$) or ($\Omega=1$, $\sigma_8=1$).
The detection of cosmic shear demonstrates the technical feasibility
of using weak lensing surveys to measure dark matter clustering
and the potential for cosmological parameter measurements,
in particular with upcoming wide field CCD cameras.

\end{abstract}


\section{Introduction}

The measurement of weak gravitational lensing produced by
the large-scale structures in the universe (hereafter, the cosmic shear)
is potentially the most effective, albeit challenging,
step toward a direct mapping of the dark matter distribution
in the universe at intermediate and low redshift. Unlike several popular probes
of large-scale structures, lensing
maps the dark matter directly, regardless of the distribution of light
emitted by gas and galaxies or the dynamical stage of the structures analysed.

A decade of theoretical and technical studies has shown that the gravitational
distortion produced by the structures along the lines-of-sight contains
important clues on structure formation models (see \cite{M99},
\cite{BS00} for reviews and references therein).  From these
studies, we know that
weak lensing can provide measurements of cosmological parameters and
the  shape of the projected density power spectrum (\cite{B91}, \cite{ME91},
\cite{K92}, \cite{V96}, \cite{B97},
\cite{JS97}, \cite{K98}, \cite{SVWJK98}, \cite{JSW99}, \cite{VW99},
\cite{BS99}). However, it is also clear that the
most challenging issues are observationals, because the 
measurement of extremely weak gravitational distortions is severely
affected by various sources of noise
and systematics such as the photon noise, the optical distortion of
astronomical telescopes and the atmospheric distortion. Therefore, the 
problem of reliable shape
measurement has also received much attention in the last few years 
(\cite{BM95},
\cite{KSB95}, \cite{VW97}, \cite{H98}, \cite{K99}, \cite{R99}, \cite{Ks99},
\cite{B00}).

Despite considerable difficulties in recovering weak lensing
signals, the potential cosmological impact of the cosmic shear
analysis has motivated several teams to
devote efforts on imaging surveys designed for the
measurement of the galaxy distortion produced by gravitational lensing, either
by observing many independent small fields, like the VLT/FORS-I
(Maoli et al in preparation),
the HST/STIS (\cite{Seitz98}), the WHT (\cite{BRE00}), or
by observing few intermediate to large fields, like
the CFHT/CFH12K-UH8K (this work and \cite{K00}), the SDSS (\cite{Annis98}) and other 
ongoing surveys.
In this paper we present the results of the analysis based
on 2 square degrees
obtained during previous independent observing runs at the CFHT with mixed I 
and V colors.
This study is part of a our weak lensing survey carried out at CFHT
(hereafter the DESCART project\footnote{http://terapix.iap.fr/Descart/})
which will cover 16 square degrees
in four colors with the CFH12K camera. Though the survey is far from
completion,  data obtained during previous runs
  have been used jointly with the first observations
  of the DESCART survey that we did in May 1999 and in November 1999
in order to demonstrate that the technical issues can be
    overcome and to better prepare the next observations.
  This set of data  permits us
already to report on the detection of a cosmic shear signal.
\\
In the following, we discuss the technique
used to extract the cosmological signal and to
measure its  amplitude and show that systematic effects are well
under control.
  The paper is organized as follows.
Section 2 describes our data sets. Section 3 discusses the details
of our PSF correction procedure and Section 4 presents the final results.
Section 5 is devoted to the discussion of
the residual systematics  and their correction. Section 6
presents a preliminary quantitative comparison of our
signal with numerical expectations of cosmological
scenarios as derived from ray-tracing simulations. Conclusions are given
in Section 7.

\section{Description of the data}

The difficulty to get a wide angle
coverage of the sky in good conditions is the reason why there is not yet
a clear detection of cosmic shear. For this work, we decided to
get the widest angular field possible, which was done at the expense of
homogeneity of the data set. However this does not impact our primary
goals which are the detection of a weak lensing signal and the test of the
control of systematics.
\\
We use in total eight different pointings mixing CFH12K and UH8K data
sets (see Table \ref{fields}). They are spread over
  five statistically independent areas, each separated by more
than 10 degrees.
The total field
covers about $6300 {\rm arcmin}^2$, and contains $3\times 10^5$
galaxies (with a number density $n_g\simeq 30$ gal/arcmin$^2$).
Note that the galaxies are weighted as discussed in Section 2.2,
and parts of the fields are masked,
so the effective number density of galaxies is about half.

\begin{table*}
\caption{List of the fields. Most of the exposures were taken in the I band
at CFHT. The total area is $1.7$ deg$^2$, and the 8 fields are uncorrelated.
}
\label{fields}
\begin{center}
\begin{tabular}{|c|c|c|c|c|c|c|c|}
\hline
Target & Name & Camera & Used area & Filter & Exp. time & Period & seeing\\
\hline
F14P1 & F1 & CFH12K & 764 ${\rm arcmin}^2$ & V& 5400 sec. & May 1999 & 
0.9"\\
F14P2 & F2 & CFH12K & 764 ${\rm arcmin}^2$ & V & 5400 sec. & May 1999 & 
0.9"\\
F14P3 & F3 & CFH12K & 764 ${\rm arcmin}^2$ & V & 5400 sec. & May 1999 & 
0.9"\\
CFDF-03 & F4 & UH8K & 669 ${\rm arcmin}^2$ & I & 17000 sec. & Dec. 1996 &
0.75"\\
SA57 & F5 & UH8K & 669 ${\rm arcmin}^2$ & I &12000 sec. & May 1998 & 0.75"\\
A1942 & F6 & UH8K & 573 ${\rm arcmin}^2$ & I & 10800 sec. & May 1998 & 
0.75"\\
F02P1 & F7 & CFH12K & 1050 ${\rm arcmin}^2$ & I & 9360 sec. & Nov. 1999 & 
0.8"\\
F02P4 & F8 & CFH12K & 1050 ${\rm arcmin}^2$ & I & 7200 sec. & Nov. 1999 & 
0.9"\\
\hline
\end{tabular}
\end{center}
\end{table*}

  All the data were obtained at the CFHT prime focus.  We used observations
spread over 4 years from 1996 to 1999, with two different cameras: the UH8K
(\cite{Lupp94}), covering a field of 28$\times$28 square arc minutes
with 0.2 arc-second per pixel  and the CFHT12K
\footnote{http://www.cfht.hawaii.edu/Instruments/Imaging/CFH12K} 
(\cite{Cuil00})
covering a field of 42$\times$28 square arc-minutes with
  0.2 arc-second per pixel as well.  Because
  these observations were initially done
for  various scientific purposes, they have been done either in
I or in V band. Table \ref{fields} summarizes the dataset.
The SA57 field was kindly provided by M. Cr\'ez\'e and A. Robin who observed
this field for another scientific
purpose (star counts and proper motions). The
  UH8K Abell 1942 data were obtained during discretionary time.
The F14 and F02 fields are part of
the deep imaging survey of 16 square-degrees in BVRI
  being conducted at CFHT jointly by several French teams.
  This survey is designed to satisfy several scientific programs, including 
the
    DESCART weak lensing program, the study of galaxy evolution
    and clustering evolution, clusters and AGN searches, and
    prepare the spectroscopic sample to be studied for the
    VLT-VIRMOS deep redshift survey  (\cite{olf98}).
CFDF-03 is one of the Canada-France-Deep-Fields (CFDF)  studied within
the framework of the Canada-France Deep Fields, with data collected with
the UH8K (Mc Cracken et al in preparation).
\\
The observations were done as usual, by splitting the total integration time 
in
individual exposures of 10 minutes each, offsetting the telescope
   by 7 to 12 arc-seconds after each image acquisition.
For the I and the V band data, we got between 7 to 13 different
  exposures per field, all with seeing conditions varying
by less than $\pm$ 0.07 arc-seconds (the others were
not co-added).
The total exposure times range from 1.75
hours in V to 5 hours in I.
\\
The total field observed  covers 2.05 square degrees,
including 0.88 square degrees in V and
1.17 square degrees in I.  However, one CCD of the UH8K and
two CCDs initially mounted on the CFH12K of the May 1999 run
have strong charge transfer efficiency problems and are not suitable
for weak lensing analysis.  Therefore, the final area
only covers
1.74 square degrees: 0.64 square degrees in
V and 1.1 square degrees in I. As we can see from Table \ref{fields}
  each field has different properties (filter, exposure time, seeing)
which makes this first data set somewhat heterogeneous.
\\
The data processing was done at the TERAPIX data center located at IAP
which has been created in order to process big images obtained
with these panoramic CCD cameras\footnote{http://terapix.iap.fr/}.
Its CPU (2 COMPAQ XP1000 with 1.2 Gb RAM memory each equipped with
DEC alpha ev6/ev67 processors) and
disk space (1.2 Tbytes) facilities permit us to handle
such a huge amount of data efficiently.
\\
For all but the CFDF-03 field, the
preparation of the detrending frames (master bias, master dark, master
flats, superflats, fringing pattern, if any) and the generation
of pre-reduced and stacked data  were
done using the FLIPS pre-reduction package
(FITS Large Image Pre-reduction software) implemented at
  CFHT and in the TERAPIX pipeline (Cuillandre et al in preparation).
In total, more than 300
Gbytes of data have been processed for this work.
\\
The CFHT prime focus wide-field corrector introduces
a large-scale geometrical distortion in the field
(\cite{Cuil96}). Re-sampling the data
over the angular size of one CCD (14 arc-minutes) cannot be
avoided if large angular offsets ($>40$ arc-seconds)
are used for the dithering pattern (like for the CFDF-03 data).
Since we kept the offsets between all individual
exposures within a 15 arc-seconds diameter disk, the
contribution of the distortion
between objects at the top and at the bottom of the CCD
between dithered exposures is kept below one tenth of
a pixel. With the seeing above 0.7 arc-second and a
sampling of 0.2 arc-second/pixel, the contribution of this
effect is totally negligible.  A simulation of the optical
distortion of the instrument shows that  the variation from one field
to another never exceeds 0.3\%, which confirms what
we expected from the CFHT optical design of the
wide-field corrector.  We discuss
this point in Section 5, in particular by confirming that the
sensitivity of the shear components with radial distances is negligible.
Also not correcting this optical distortion results
in a slightly different plate scale from the center
to the edge of the field (pixels see more sky in the
outside field). But this is also of no consequence
for our program since the effect is very small as compared to 
 the signal we are interested in.
\\
The stacking  of the
non-CFDF images has been done independently for each individual CCD
(each covering 7$\times$14 arc-minutes).
  We decided not to create a single large UH8K or CFH12K image
per pointing since it
is useless for our purpose. It complicates the weak lensing
analysis, in particular for the PSF correction, and needs to
  handle properly the gaps between CCDs which potentially could  produce
discontinuities in the  properties of the field.  The drawback is
that we restricted ourselves to a weak lensing analysis on scales smaller 
than 7 arc-minutes
(radius smaller than 3.5 arc-minutes, as shown in the
next figures);
this is not a critical scientific issue since the total field of view
is still too
small to provide significant signal beyond that angular scale.
In the following we consider each individual CCD as
one unit of the data set.
\\
The co-addition was performed by computing first
the offset of each CCD between each individual exposure from the
identification of
common bright objects (usually  20 objects)
  spread over one of the CCD's arbitrary chosen as a reference frame.
Then, for each exposure the offsets in the x- and y- directions are
computed using the detection algorithm of the
SExtractor package (\cite{BA96}) which provides a
typical accuracy better than one tenth
of a pixel for bright objects.  The internal accuracy of this
technique is given by the rms fluctuations of the offsets of each
reference object. Because our offsets were small the procedure works
very well and provides a stable solution quickly. We usually reach
  an accuracy over the CCDs
of 0.25 pixels rms (0.05 arc-second) in both directions
for offsets of about 10 arc-seconds (50 pixels).  Once the offsets
are known the individual CCDs are stacked  using
  a bilinear interpolation
  and by oversampling each pixel by a factor of 5
in both x- and y- directions  (corresponding to
the rms accuracy of the offsets). The images are then re-binned 1$\times$1
  and finally a clipped median
procedure is used for the addition.
The procedure requires
CPU and disk space but works very well, provided the shift between
exposures remains small. We then end up with a final set
of stacked CCDs which are ready for weak lensing analysis.
\\
The  twelve separate
pointings of the  CFDF-03 field were processed independently
using
a method which is fully described elsewhere (McCracken et al 2000, in
preparation). Briefly, it uses astrometric sources present in
the field to derive a world coordinate system (WCS; in this work we
use a gnomic projection with higher order terms). This mapping is then
used to combine the eight CCD frames to produce a single image in which
a uniform pixel scale is restored across the field. Subsequent
pointings are registered to this initial WCS by using a large number of
sources distributed over the eight CCDs to correct for telescope
flexure and atmospheric refraction. For each pointing the registration
accuracy is $\sim 0.05''$ rms over the entire field. The final twelve
projected images are combined using a clipped median, which, although
sub-optimal in S/N terms, provides the best rejection for cosmic rays
and other transient events for small numbers of input images.

\section{Galaxy shape analysis}

The galaxies have been
  processed using the IMCAT software generously made available by Nick 
Kaiser\footnote{http://www.ifa.hawaii.edu/$\sim$kaiser/}. Some of 
the process steps
have been modified from the original IMCAT version
in order to comply with our specific needs. These modifications are 
described now.

The object detection, centroid, size and magnitude measurements
are done using SExtractor
(\cite{BA96}\footnote{see also ftp://geveor.iap.fr/pub/sextractor/})
which is optimized for the detection of galaxies.
We replace the
  parameter $r_g$ (physical size of an object), calculated
in the IMCAT peak finder algorithm
by the half-light-radius of
SExtractor (which is very similar to $r_h$ measured in IMCAT).
This lowers the signal-to-noise of the shape measurements slightly,
but it is not a serious issue for the statistical detection of cosmic shear
described in this work.
Before going into the details of the shape analysis, we first
briefly review how IMCAT measures shapes and corrects
the stellar anisotropy. Technical details and proofs can be found in
\cite{KSB95} (hereafter KSB), \cite{H98} and \cite{BS00}.

\subsection{PSF correction: the principle}

KSB derived how a gravitational shear and an anisotropic PSF
affect the shape of a galaxy. Their
derivation is a first order effect calculation, which has the nice
property to separate the gravitational shear and the atmospheric effects. The
correction is calculated first on the second moments of a galaxy, and 
subsequently the galaxy
ellipticity can be directly expressed as a function of the shear and
the star anisotropy. The raw ellipticity
$\eg$ of an object is the
quantity measured from the second moments $I_{ij}$ of the surface brightness
$f(\thetag)$:

\begin{equation}
\eg=\left({I_{11}-I_{22}\over Tr(I)} ; {2I_{12}\over Tr(I)}\right), \ \ \
I_{ij}=\int {\rm d}^2\theta
W(\theta)\theta_i\theta_j f(\thetag).
\end{equation}
The aim of the window function $W(\theta)$ is to
suppress the photon noise which dominates
the objects profile at large radii.
According to KSB, in the presence of a shear $\gamma_\beta$ and a PSF anisotropy $p_\beta$,
the raw ellipticity $\eg$ is
sheared and smeared, and modified by the quantity $\delta \eg$:

\begin{equation}
\delta e_\alpha=P^{sh}_{\alpha\beta}\gamma_\beta+P^{sm}_{\alpha\beta} 
p_\beta.
\label{eshift}
\end{equation}
The shear and smear polarization tensors $P^{sh}_{\alpha\beta}$ and
$P^{sm}_{\alpha\beta}$ are
measured from the data, and the stellar ellipticity $\pg$, also measured 
from
the data,
is given by the raw stellar ellipticity $\eg^\star$:

\begin{equation}
p_\alpha={e_\alpha^\star\over P^{sm}_{\alpha\alpha}}.
\label{estar}
\end{equation}
Using Eq.(\ref{eshift}) and Eq.(\ref{estar}) we can therefore correct
for the stellar anisotropy, and
obtain an unbiased estimate of the orientation of the shear $\gamma_\beta$. 
To
get the right amplitude
of the shear, a piece is still missing: the isotropic correction, caused by 
the
filter $W(\theta)$
and the isotropic part of the PSF, which tend to circularize
the objects. \cite{LK97} absorbed this isotropic correction
by replacing the shear polarization $P^{sh}$ in Eq(\ref{eshift}) (which is
an exact derivation in the case
of a Gaussian PSF) by
the pre-seeing shear polarisability $P^\gamma$:

\begin{equation}
P^\gamma=P^{sh}-{P^{sh}_\star\over P^{sm}_\star} P^{sm}.
\end{equation}
This factor 'rescales' the galaxy ellipticity to its true value without changing
its orientation, after the stellar anisotropy term was removed. The residual
anisotropy left afterwards is the cosmic shear $\gamma_\beta$, therefore
the observed ellipticity can be written as the sum of a 'source' 
ellipticity,
a gravitational shear $\gammag$ term, and a stellar anisotropy contribution:

\begin{equation}
\eg^{obs}=\eg^{source}+P_\gamma\gammag+P^{sm} \pg.
\label{imcateq}
\end{equation}
There is no reason that $\eg^{source}$ should be the {\it true} source
ellipticity $\eg^{true}$, as
demonstrated by \cite{BS00}. The only thing we know about $\eg^{source}$ is 
that $\langle\eg^{true}\rangle =0$ implies $\langle\eg^{source}\rangle =0$.
Therefore Eq.(\ref{imcateq})
provides an unbiased estimate of the shear $\gammag$ as long as the 
intrinsic ellipticities of the galaxies are
uncorrelated (which leads to $\langle\eg^{true}\rangle =0$).
The estimate of the shear is simply given by

\begin{equation}
\gamma=P_\gamma^{-1}\cdot(\eg^{obs}-P^{sm} \pg) .
\label{shearestim}
\end{equation}

The quantities $P^\gamma$, $P^{sm}$ and $\pg$ are calculated for each 
object.
The shear estimate per galaxy (Eq(\ref{shearestim})) is done using the 
matrices of the
different polarization tensors, and not their traces (which
corresponds to a scalar correction) as often done
in the literature.
Although the difference between tensor and scalar correction is small 
(because
$P_\gamma$ is
nearly proportional to the identity matrix), we show elsewhere,
in a comprehensive simulation paper
(\cite{E00}), that the tensor correction
gives slightly better results.

\subsection{PSF correction: the method}

The process of galaxy detection and shape correction can be done 
automatically, provided we first
have a sample of stars representative of the PSF.  However,
in practice the star selection needs careful attention and cannot be
automated because of contaminations.  Stars
can have very close neighbor(s)
(for instance a small galaxy exactly aligned with it) that their
shape parameters are strongly affected.
Therefore we adopted a slow but well-controlled manual star selection
process: on each
  CCD, the stars are first selected in  the
stellar branch of the $r_g-mag$ diagram in order to be certain to
eliminate saturated and very
faint stars. We then perform a $3\sigma$ clipping on the {\it corrected} 
star
ellipticities, which
removes most of the stars whose shape is affected by
neighbors (they behave as outliers compared to the surounding
stars). It is worth noting that the $\sigma$ clipping
should be done on the
corrected ellipticities and not on the raw ellipticities, since only the
corrected ellipticities
are supposed to have a vanishing anisotropy.
The stellar outliers which survived the $\sigma$ clipping are checked by eye
individually
to make sure that no unusual systematics are present.\\
During this procedure,
we also manually mask the regions of the CCD which could potentially
produce artificial signal.  This includes for example
the areas  with very strong gradient of the sky background,
like  around bright stars or bright/extended galaxies, but also
  spikes produced
along the diffraction image of the spider supporting the secondary mirror,
   columns containing light from saturated stars,
  CCD columns with bad charge transfer efficiency, residuals from
transient events like asteroids
which cross the CCD during the  exposure and finally all the boundaries of
each CCD.  At the end, we are left with a raw galaxy catalogue and a star
catalogue free of spurious objects, and each CCD chip has been checked
individually. This masking process removes about 15\% of the CCD area
and the selection itself leaves about 30 to 100 usable stars per CCD.

The most difficult step in the PSF correction is Eq.(\ref{shearestim}), 
where the
inverse of a noisy matrix $P^\gamma$
is involved. If we do not pay attention to this problem, we obtain corrected
ellipticities which can be very large
and/or negative, which would force us to apply severe cuts on the final 
catalogue to
remove aberrant corrections, thus losing many objects. A natural
way to solve the problem is to smooth the
matrix $P^\gamma$ before it is inverted.  In principle
$P^\gamma$ should be
smoothed in the largest possible parameter space defining
the objects: $P^\gamma$ might
depend on the magnitude, the ellipticity, the profile, the size,
etc...  In practice, it is common to smooth
$P^\gamma$ according to the magnitude and the size (see for instance
\cite{Ketal98}, \cite{H98}). Smoothing performed on a regular grid
is generally not optimal, and instead, we calculate the smoothed $P^\gamma$ for each
galaxy from its nearest neighbors in the objects parameter space (this has 
the advantage of finding locally the optimal mesh size for grid smoothing).
Increasing the parameter space for smoothing does not lead to significant
improvement in the correction (which is confirmed by our simulations in \cite{E00}),
therefore we keep the magnitude and the size $r_h$ to be the main functional
dependencies of $P^\gamma$.

A smoothed $P^\gamma$ does not eliminate all abnormal ellipticities; the
next step is
to weight the galaxies according to the noise level of the ellipticity
correction. Again, this can be done in
the gridded magnitude/size parameter space where each cell contains a fixed
number of objects (the nearest
neighbors method). We then calculate the
variance $\sigma^2_\epsilon$ of the ellipticity of those galaxies, which 
gives
an indication of the dispersion
of the ellipticities of the objects in the cell: the
larger $\sigma^2_\epsilon$, the larger the noise. We then calculate
a {\it weight} $w$ for each galaxy, which is directly given by
$\sigma^2_\epsilon$:

\begin{equation}
w=\Big\{\matrix{ \exp(-5(\sigma_\epsilon-\alpha)^2)~~{\rm if~\sigma_\epsilon 
<1}
\cr \displaystyle{{1\over \sigma^2_\epsilon}} \exp(-5(1-\alpha)^2) ~~{\rm
if~\sigma_\epsilon >1}},
\label{weightfct}
\end{equation}
where $\alpha$ is a free parameter, which is chosen to be the maximum of the
ellipticity distribution of the galaxies.
Eq.(\ref{weightfct}) might seem arbitrary compared to the usual
$1/\sigma^2_\epsilon$ weighting, but
the inverse square weighting tends to diverge for
low-noise objects (because such objects have a small
$\sigma_\epsilon^2$), which create a strong unbalance among low noise objects.
The aim of the exponential cut-off as defined in Eq.(\ref{weightfct})
is to suppress this divergence\footnote{Note that the use of a different weighting 
scheme like $w\propto 1/(\alpha^2+
\sigma_\epsilon^2)$ has almost no effect on the detection. Other weighting schemes
have been used, such as in \cite{HFK00}}.

The weighting function prevents the use of an arbitrary and sharp cut
to remove the bad objects. However, we found in our simulations
(\cite{E00}) that we should remove
objects smaller than the seeing size, since they carry very little
  lensing information, and the PSF convolution
is likely to dominate the shear amplitude. Our final catalogue contains 
about
$191000$ galaxies, of which
$23000$ are masked. It is a galaxy number density of about $n\simeq 26~{\rm
gal/arcmin}^2$, although the
effective number density when the weighting is considered should be much 
less.
We find
$\alpha=0.5$, which corresponds to the ellipticity variance of the whole
catalogue.

\section{Measured signal}

The quantity directly accessible from the galaxy shapes and related to
the cosmological model
is the variance of the shear $\langle\gamma^2\rangle$. An analytical 
estimate of
it using a
simplified cosmological model (power-law power spectrum, sources at a single
redshift plane,
leading order of the perturbation theory, and no cosmological constant) 
gives
(\cite{K92}, \cite{V96},
\cite{B97}, \cite{JS97}):

\begin{equation}
\langle\gamma^2\rangle^{1/2}\simeq 0.01\sigma_8 \Omega_0^{0.75} z_s^{0.75}
\left({\theta \over 1{\rm arcmin}}\right)^{-\left({n+2\over 2}\right)},
\label{vartheo}
\end{equation}
where $n$ is the slope of the power spectrum, $\sigma_8$ its normalization,
$z_s$ the redshift of the sources and
$\theta$ the top-hat smoothing filter radius. The
expected effect is at the percent level, but at small scales
  the non-linear dynamics is expected to increase the
signal by a factor of a few (\cite{JS97}). Nevertheless Eq.(\ref{vartheo}) 
has
the advantage of clearly giving
the cosmological dependence of the variance of the shear.

\begin{figure}
\centerline{
\psfig{figure=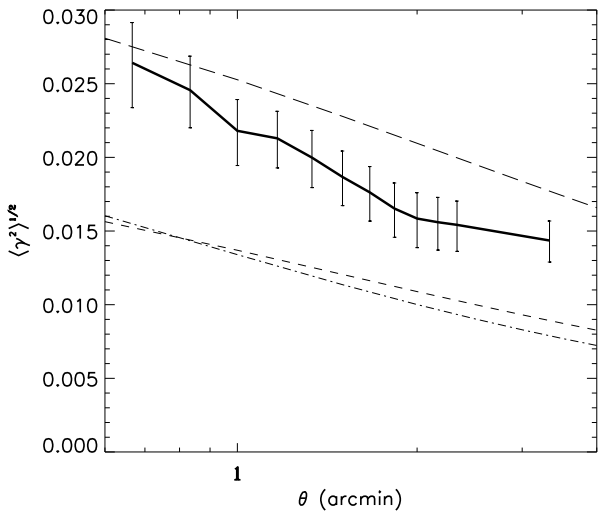,height=7cm}}
\caption{\label{9700.f1.ps} Square-root of the variance of the measured
shear as a function of the radius of the top-hat window (solid line).
The maximum angular scale, 3.5 arc-minutes radius, is fixed by the maximum
angular scale defined by individual CCDs (7'). Error
bars are computed over $1000$ random
realizations of the galaxy catalogue. The other lines are theoretical
  predictions
of the same quantity for different cosmological models in the non-linear 
regime
(using the fitting formula in \cite{PD96}):
the long-dashed line corresponds to $(\Omega=1, \Lambda=0, \sigma_8=0.6)$, 
the
dashed line to $(\Omega=0.3, \Lambda=0, \sigma_8=0.6)$, and the dot-dashed 
line
to $(\Omega=0.3, \Lambda=0.7, \sigma_8=0.6)$.}
\end{figure}

\begin{figure*}
\centerline{
\psfig{figure=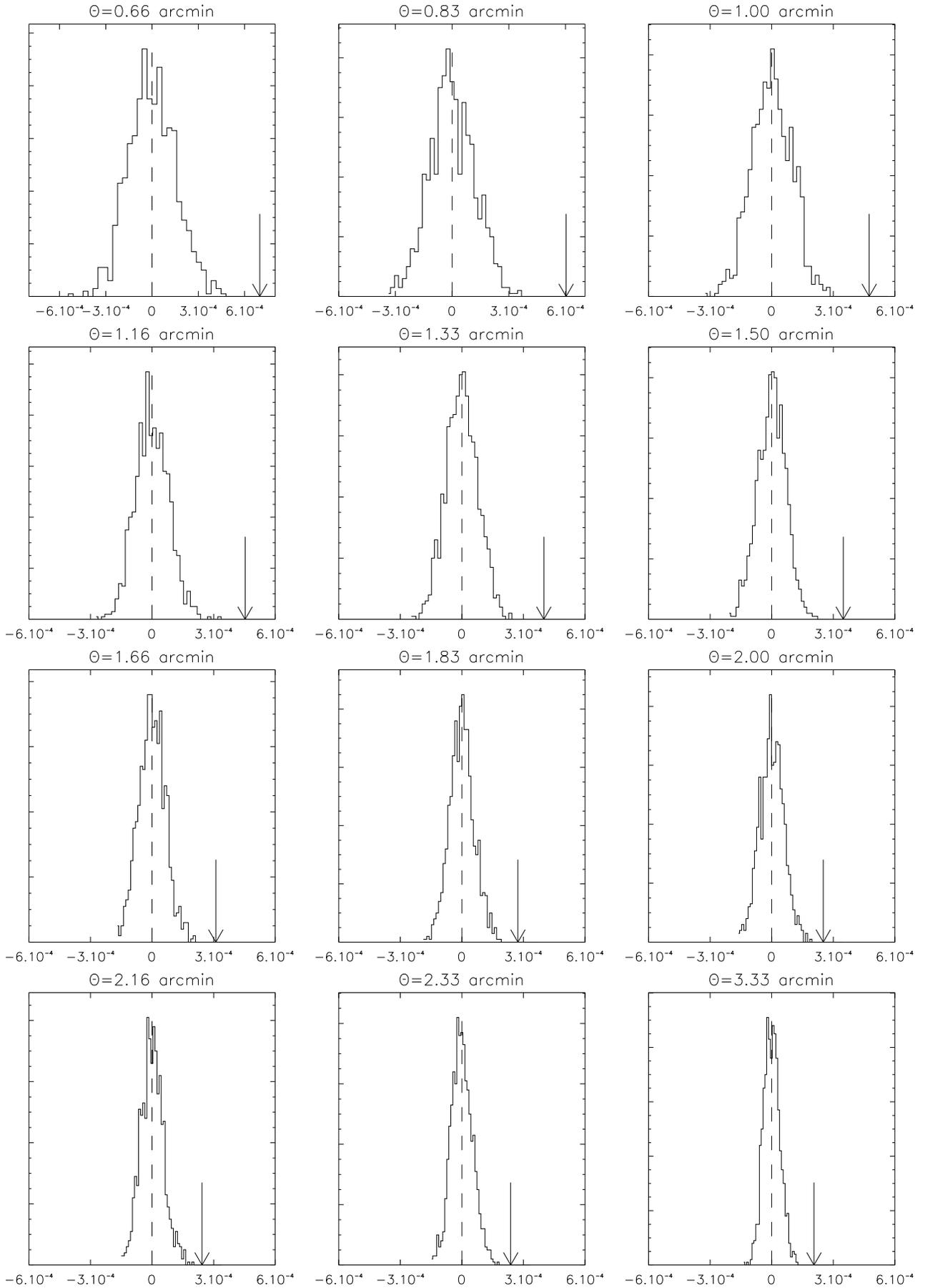,height=24cm}}
\caption{\label{9700.f2.ps} For different smoothing sizes (indicated at
the top of each panel), the value of the measured signal (given by the 
arrow)
compared to the signal measured in the randomized catalogues (histograms). 
This
figure shows how far the signal deviates from a pure random orientation of the
galaxies.
Note that the distribution of $\langle \gamma^2\rangle$ is not 
Gaussian.
}
\end{figure*}

From the unweighted galaxy ellipticities $e_\alpha$, an estimate of
$\gamma^2(\thetag_i)$ at the position
$\thetag_i$ is given by:

\begin{equation}
E[\gamma^2(\thetag_i)]={\displaystyle \sum_{\alpha=1,2}} \left({1\over
N}{\displaystyle \sum_{k=1}^N} e_\alpha
(\theta_k)\right)^2.
\label{estimator}
\end{equation}
The inner summation is performed over the $N$ galaxies located inside the
smoothing window centered on
$\thetag_i$, and the outer summation over the ellipticity components. The
ensemble average of
Eq.(\ref{estimator}) is

\begin{equation}
\langle E[\gamma^2(\thetag_i)]\rangle={\sigma_\epsilon^2\over
N}+\langle\gamma^2\rangle.
\end{equation}
The term $\sigma_\epsilon^2/N$ can be easily removed using a random
  realization of the galaxy catalogue:
each position angle of the galaxies is randomized, and the variance of the 
shear
is calculated again. This randomization
  allows us to determine $\sigma_\epsilon^2/N$ and the error bars associated
with the noise due to the
intrinsic ellipticity distribution. At least 1000 random
realizations are required in order to have
a precise estimate of the error bars. Note that it is strictly equivalent to 
use
an estimator where the diagonal terms are
removed in the sum (\ref{estimator}), which suppress automatically the
$\sigma_\epsilon^2/N$ bias.

When we take into account the weighting scheme for each galaxy,
the estimator Eq.(\ref{estimator}) has to be modified accordingly
as follows:

\begin{equation}
E[\gamma^2(\thetag_i)]={\displaystyle \sum_{\alpha=1,2}} 
\left({{\displaystyle
\sum_{k=1}^N} w(\theta_k) e_\alpha
(\theta_k)\over {\displaystyle \sum_{k=1}^N} w(\theta_k)} \right)^2,
\label{theestimator}
\end{equation}
where $w$ is the weight as defined in Eq.(\ref{weightfct}).
The variance of the shear is not only the easiest quantity to measure, but 
it is also fairly weakly
sensitive to the systematics provided that they are smaller than the signal.
The reason is that any spurious
alignment of the galaxies, in addition to the
gravitational effect, adds quadratically to the signal and not linearly:

\begin{equation}
\langle \gamma^2_{mes}\rangle=\langle \gamma^2_{true}\rangle+\langle
\gamma^2_{bias}\rangle.
\end{equation}
Therefore, a systematic of say $1\%$ for a signal of $3\%$ only contributes 
to
$\sim 5\%$ in $\langle\gamma^2
\rangle^{1/2}$. We investigate in detail in the next Sections the term 
$\langle
\gamma^2_{bias}\rangle$ and show that
it has a negligible contribution.

We will present results on the shear variance measured from the
data sets described in Section 2. The variance $\langle 
\gamma^2_{mes}\rangle$ is measured in apertures which are placed on
a $10\times 20$ grid for each of the $2000\times 4000$ CCDs. By construction
the apertures never cross the CCD boundaries, and if more than $10\%$ of
the included objects turns out to be masked objects, this aperture is not used.
Figure \ref{9700.f1.ps} shows $\langle \gamma^2_{mes}\rangle^{1/2}$ 
(thick
line) with error bars obtained from
$1000$ random realizations. The three other thin lines correspond to 
theoretical
predictions obtained from an
exact numerical computation for three different cosmological models, in the
non-linear regime. We assumed a
normalized broad source redshift distribution given by

\begin{equation}
n(z_s)={\beta\over z_0 \Gamma\left({1+\alpha\over \beta}\right)} 
\left({z_s\over
z_0}\right)^\alpha \exp\left(-\left({z_s\over z_0}\right)^\beta\right),
\end{equation}
with the parameters $(z_0,\alpha,\beta)=(0.9,2,1.5)$ are supposed to
match roughly the redshift distribution in our
data sets\footnote{with a source redshift distribution which peaks
at 0.9}.  The shape of this redshift distribution mimics those observed
in spectroscopic magnitude-limited samples as well as those inferred from
theoretical predictions of galaxy evolution models.
Since we did significant selections in our galaxy catalog the final
redshift distribution could be modified. We have not quantified
this, but we do not think it would significantly change the average
redshift of the sample, even if the shape may be modified. The
variance of the shear $\langle\gamma^2\rangle$ is computed via the formula 
(see
\cite{SVWJK98} for
the notations):

\begin{equation}
\langle\gamma^2\rangle=2\pi \int_0^\infty~k{\rm d}k P_\kappa(k)
I^2_{TH}(k\theta),
\end{equation}
where $I^2_{TH}$ is the Fourier transform of a Top-Hat window function, and
$P_\kappa(k)$ is the
convergence power spectrum, which depends on the projected 3-dimensional 
mass
power spectrum $P_{3D}(k)$:

\begin{eqnarray}
P_\kappa(k)&=&{9\over 4}\Omega_0^2\int_0^{w_H} {{\rm d}w \over a^2(w)}
P_{3D}\left({k\over f_K(w)};
w\right)\times\nonumber\\
&&\int_w^{w_H}{\rm d} w' n(w') {f_K(w'-w)\over f_K(w')}.
\end{eqnarray}
$f_K(w)$ is the comoving angular diameter distance out to a distance $w$
($w_H$ is the horizon distance),
and $n(w(z))$ is the redshift distribution of the sources. The nonlinear 
mass
power spectrum $P_{3D}(k)$
is calculated using a fitting formula (\cite{PD96}).

We see in Figure \ref{9700.f1.ps} that the measured signal
is consistent with the theoretical prediction,
both in amplitude and in shape. In order to have a better
idea of how  significant the signal is we can compare
for each smoothing scale the histogram of the shear variance in the
randomized samples and the measured signal. This
is is shown in Figure \ref{9700.f2.ps}, for all the smoothing scales
shown in Figure \ref{9700.f1.ps}.
The signal is significant up to a level of $5.5\sigma$ . Note that the measurement
points at different scales are correlated, and that an estimate of the overall
significance of our signal would require the computation of the noise correlation
matrix between the various scales.

\section{Analysis of the systematics}

Now we have to check that the known systematics cannot be responsible for the
signal. In the following we discuss three
types of systematics:

\begin{itemize}
\item[$\bullet$] The intrinsic alignment of galaxies which could exists
in addition to the
lensing effect. We assume such an alignment do not exist, but the
overlapping isophotes of close galaxies produces it. We could in 
principle
remove this effect by choosing a window function
small compared to the galaxy distance in the pair, such that close galaxies 
do
not influence the second moment
calculation of themselves. However this is difficult to achieve in practice.

\item[$\bullet$] The strongest known systematics is the PSF anisotropy 
caused by
telescope tracking errors,
the optical distortion, or any imaginable source of anisotropy of the
star ellipticity. We have to be sure that
the PSF correction outlined in Section 2.2 removes any correlation between
galaxy and star ellipticities.

\item[$\bullet$] The spurious alignment of galaxies along the CCD frame
lines/columns. We cannot reject
this possibility since charge transfer along the readout directions
is done by moving the charges
  from one pixel to the next pixel and so forth,
with an transfer efficiency of $0.99998$. This  effect
could spread the charges of the bright objects (very bright and
saturated stars produce
this kind of alignment, but they have been removed during the
masking procedure).
   Therefore we can expect the objects to be elongated along the readout
direction.
\end{itemize}

\subsection{Systematics due to overlapping isophotes}

\begin{figure}
\centerline{
\psfig{figure=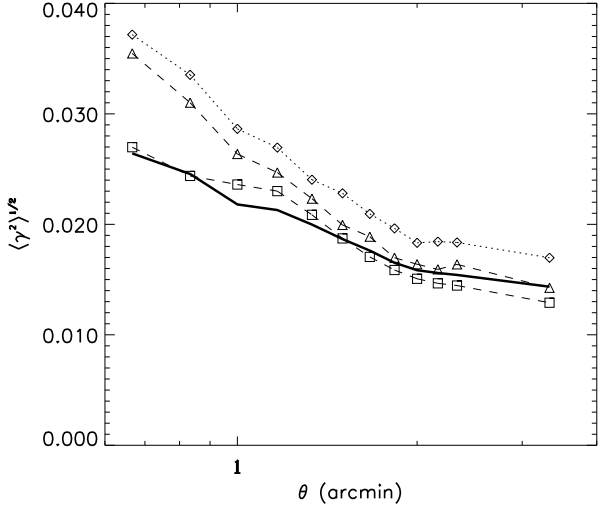,height=7cm}}
\caption{\label{9700.f3.ps} The thick solid line shows the signal as plotted 
on
Figure \ref{9700.f1.ps}. It was obtained with a catalogue of galaxies 
where
galaxies closer than $10$ pixels were rejected. The three other curves show
the same signal measured with different rejection criteria: the 
diamond-dotted
line
is for no rejection at all, the triangle-dashed line for galaxies closer 
than
$5$ pixels rejected, and the square-dashed line for galaxies closer than 
$20$
pixels
rejected. This figure illustrates  that the overlapping isophotes of
close galaxies tends to overestimate the shear.}
\end{figure}

Let us consider the first point in the above list of systematics. In order 
to
study the effect of close galaxy
pairs, we measured the signal by removing close pairs by varying a cut-off 
applied on
the respective distance of close galaxies.
Figure \ref{9700.f3.ps} shows the signal measured when successively closer 
pairs
with $d=0$ (no pair rejection),
$5$, $10$ and $20$ pixels have been rejected. The cases $d=0$ and $d=5$ show 
an
excess of power at small scales
compared to $d=10$ and $d=20$ (the latter two give the same signal). 
Therefore we
assume that for $d>10$ we have
suppressed the overlapping isophote problem, and in the following we keep 
the
$d=10$ distance cut-off, which
gives us a total of $\sim 168000$ galaxies for the whole data sets, as 
already
indicated at the end of
Section 3.2. By removing close
pairs of galaxies, we also remove the effect of possible alignment of the 
ellipticities of
galaxies in a group due to tidal forces.

\subsection{Systematics due to the anisotropic PSF correction}

\begin{figure}
\centerline{
\psfig{figure=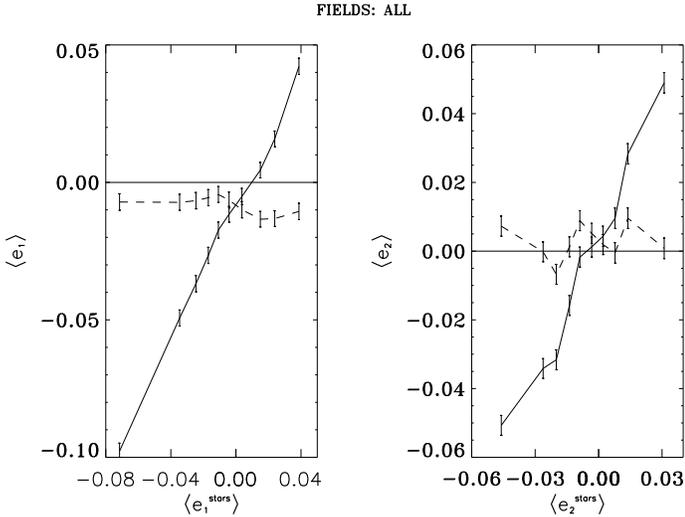,height=7cm}}
\caption{\label{9700.f4.ps} Average galaxy ellipticity $\langle
e_\alpha\rangle$
versus the average star ellipticity $\langle e^\star_\alpha\rangle$ for
both components $\alpha=1,2$. The dashed lines are obtained from the fully
corrected galaxy ellipticities, as given by Eq.(\ref{imcateq}). The solid
lines are obtained
from the galaxy ellipticities corrected from
the seeing, but without the anisotropy correction
term $P^{\rm sm}\pg$ of  Eq.(\ref{imcateq}). Each ellipticity
bin contains about $N=16000$ galaxies, and the error bars are calculated
assuming
Gaussian errors $\propto N$. Except for a constant tiny bias along the $e_1$
direction, the corrected galaxies are uncorrelated with the stellar 
ellipticity,
which
demonstrates that the PSF correction method works well.
}
\end{figure}

\begin{figure}
\centerline{
\psfig{figure=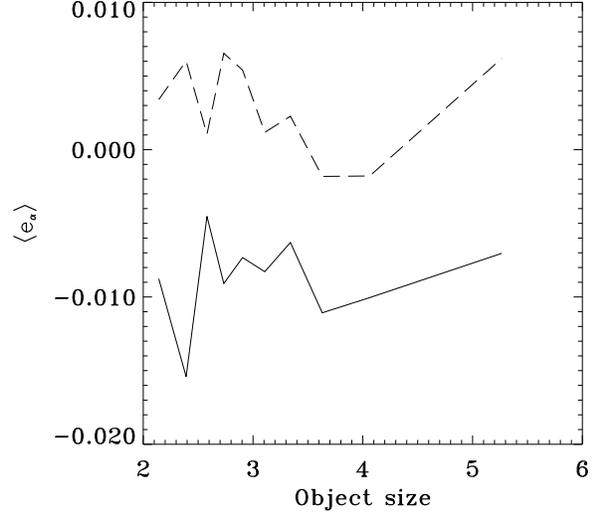,height=7cm}}
\caption{\label{9700.f5.ps} Average galaxy ellipticity $\langle e_1\rangle$ (solid line)
and $\langle e_2\rangle$ (dashed line) as a function of the object size $r_h$. It is shown that
the systematic bias of $-1\%$ along the $e_1$ component is fairly galaxy independent.
}
\end{figure}

\begin{figure}
\centerline{
\psfig{figure=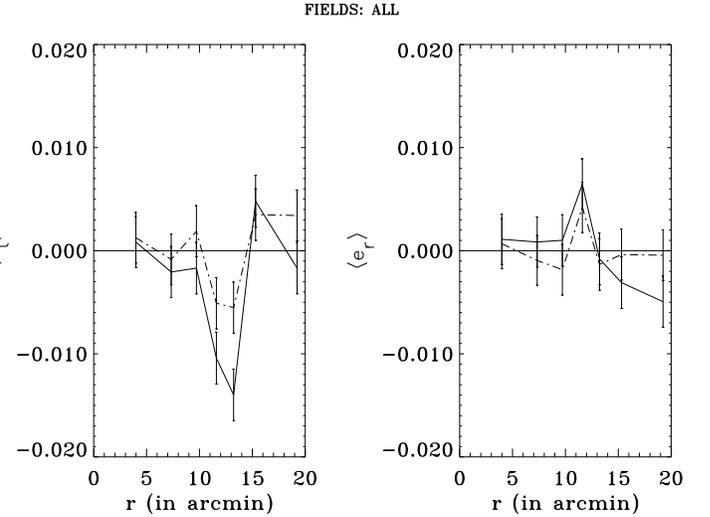,height=7cm}}
\caption{\label{9700.f6.ps} Average tangential galaxy ellipticity
$\langle \gamma_t\rangle$ and radial galaxy ellipticity
$\langle \gamma_r\rangle$ versus the distance from the optical center $r$.
As for Figure \ref{9700.f4.ps}, the dashed lines are obtained from the 
 fully corrected 
 galaxy ellipticities, as given by Eq.(\ref{imcateq}), and the 
solid
lines are obtained
from  galaxy ellipticities corrected from seeing but  
  where the anisotropy correction
term $P^{\rm sm}\pg$ has been removed from Eq.(\ref{imcateq}). Each 
ellipticity
bin contains about $N=24000$ galaxies, and the error bars are calculated
assuming
Gaussian errors $\propto N$. The absence of a significant amplitude between
the dashed and the solid lines show that the optical distortion effect is a
negligible contribution to the PSF anisotropy.
}
\end{figure}

\begin{figure*}
\psfig{figure=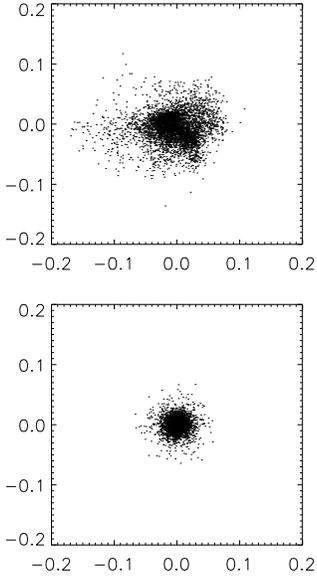,height=8cm}
\caption{\label{9700.f11.ps} Star ellipticities of all the survey before
(top panel) and after (bottom panel) the correction. After correction,
the star ellipticity is randomly distributed around zero, as expected.}
\end{figure*}

\begin{figure*}
\centerline{
\psfig{figure=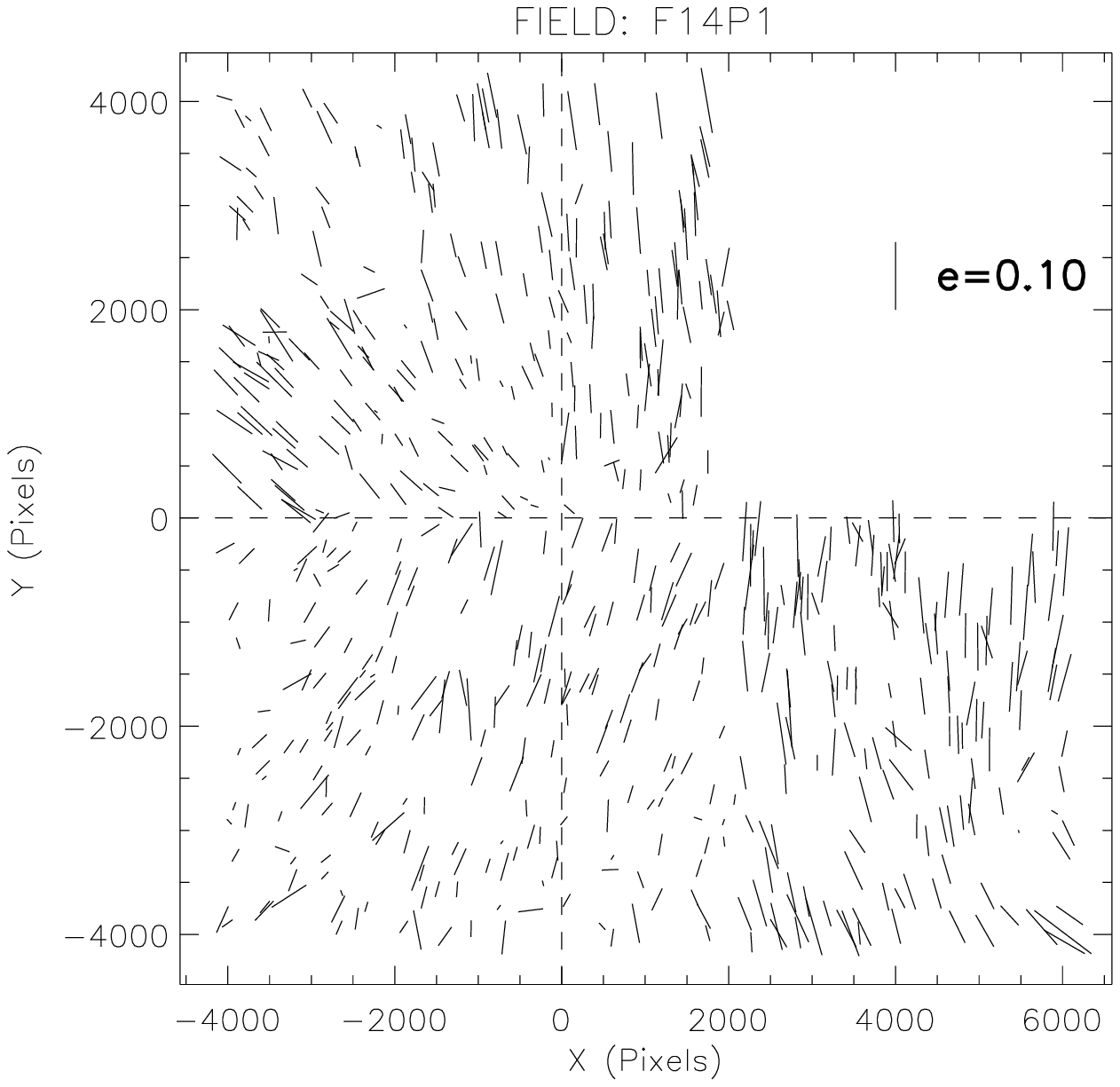,height=8cm}
\psfig{figure=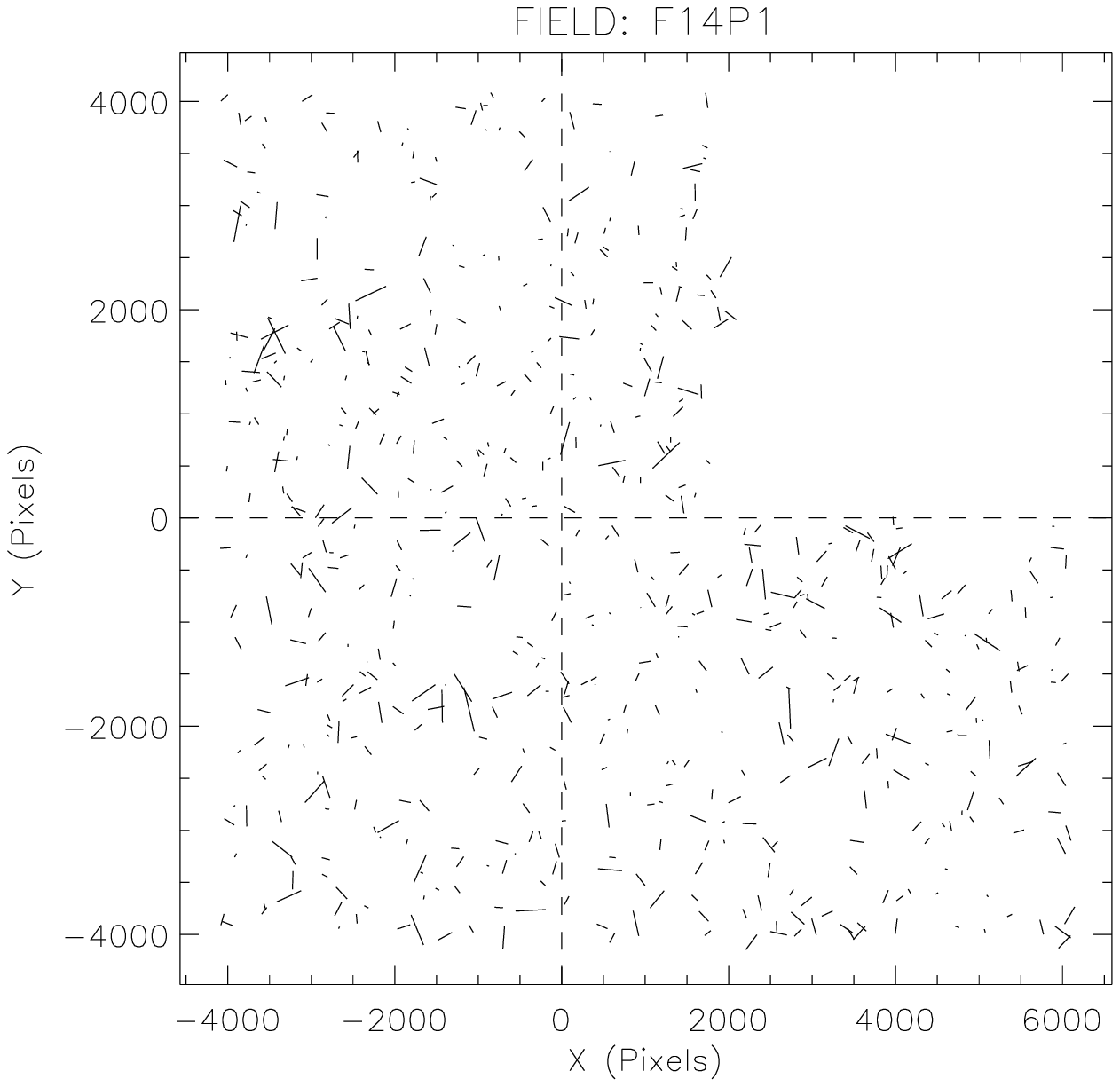,height=8cm}}
\caption{\label{F14P1.ps} Uncorrected (left) and corrected (right) star
ellipticities for FIELD F14P1. The dashed
cross shows the location of the optical center. Frames are graduated in 
pixels. The reference stick at the top-left of the frame shows the amplitude
of a $10\%$ distortion. This length of reference applies also for Figures \ref{F14P2.ps} to
\ref{F02P4.ps}.}
\end{figure*}

\begin{figure*}
\centerline{
\psfig{figure=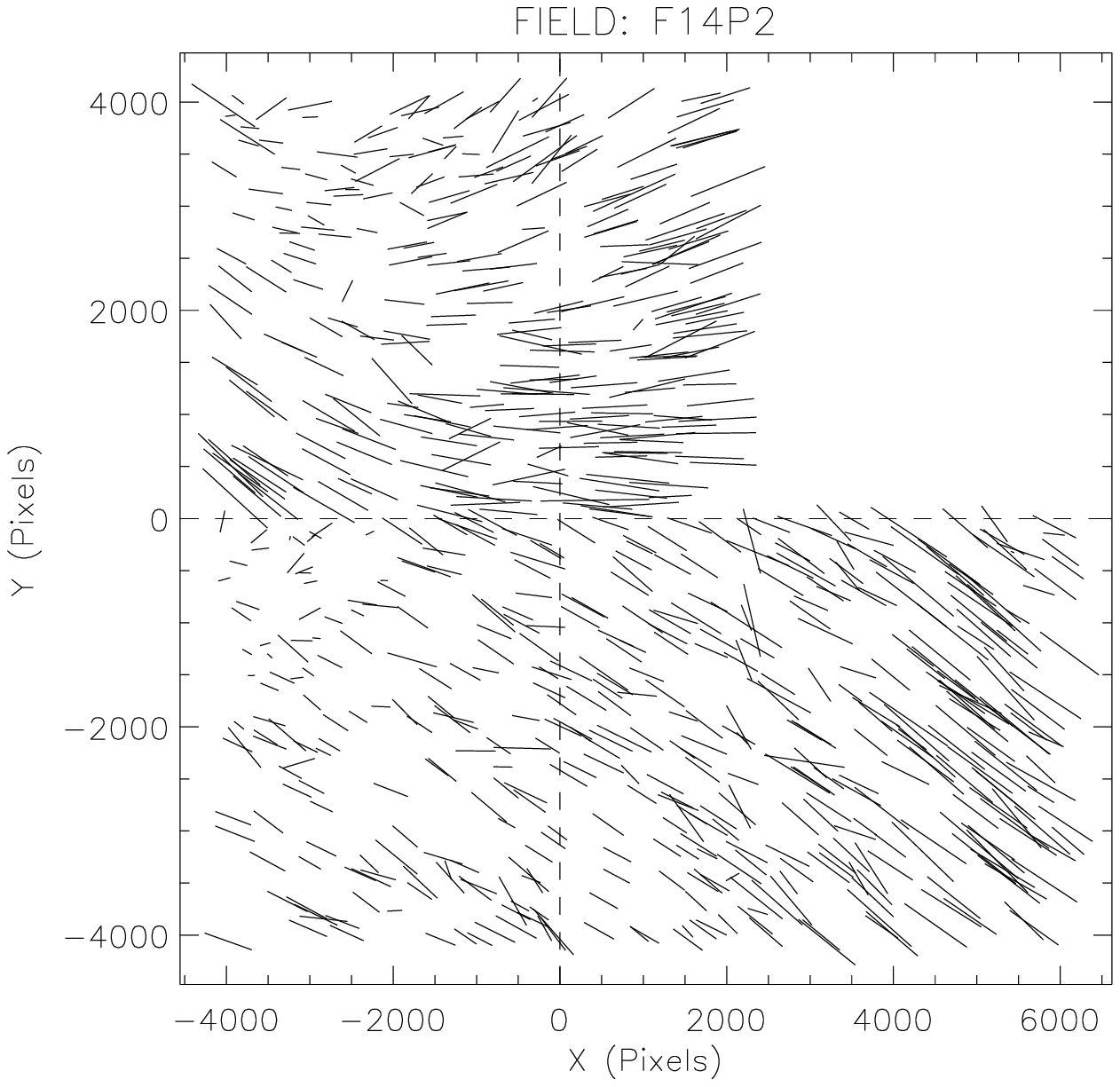,height=8cm}
\psfig{figure=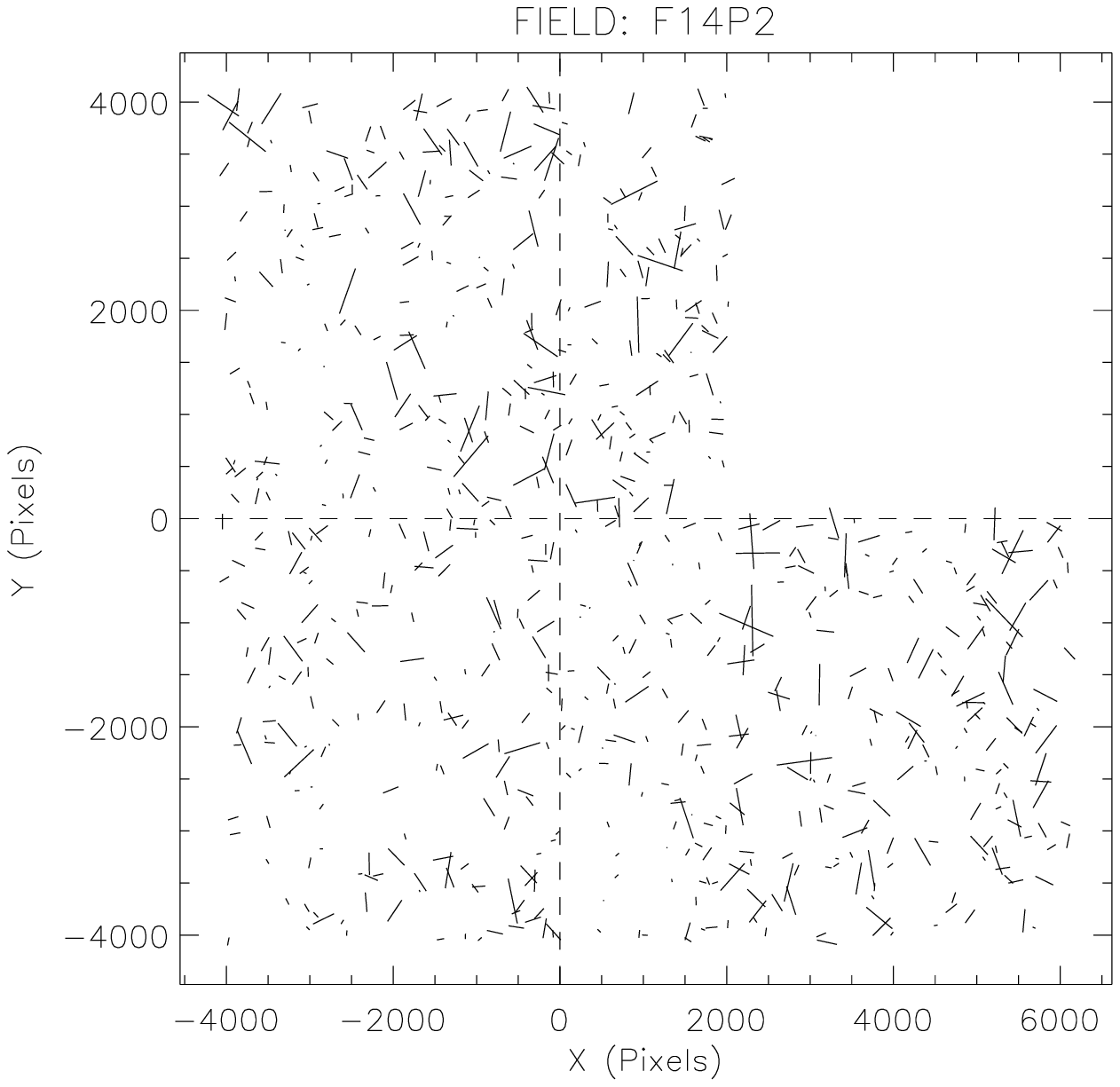,height=8cm}}
\caption{\label{F14P2.ps} Same as Figure \ref{F14P1.ps} for FIELD F14P2.}
\end{figure*}

\begin{figure*}
\centerline{
\psfig{figure=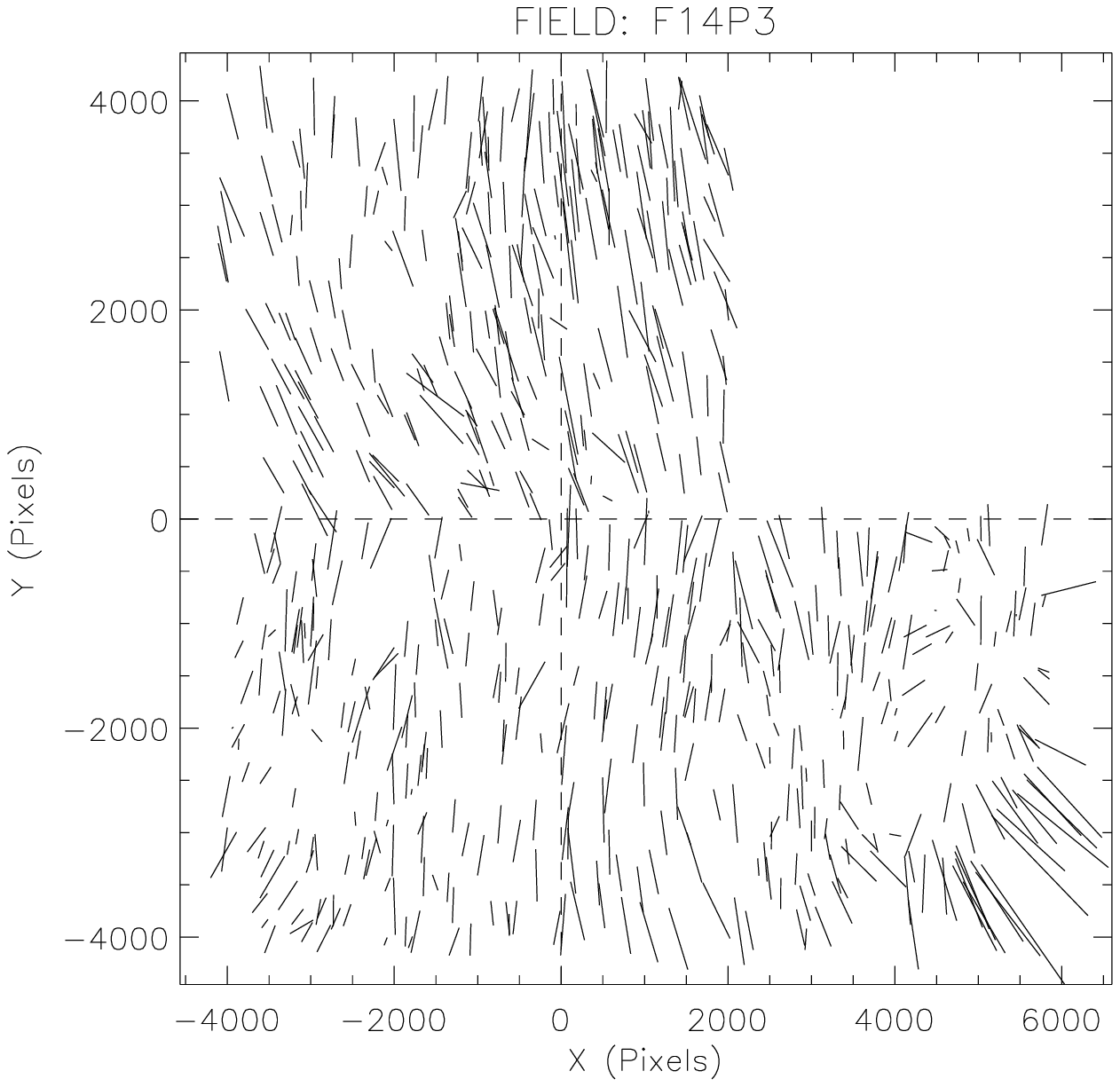,height=8cm}
\psfig{figure=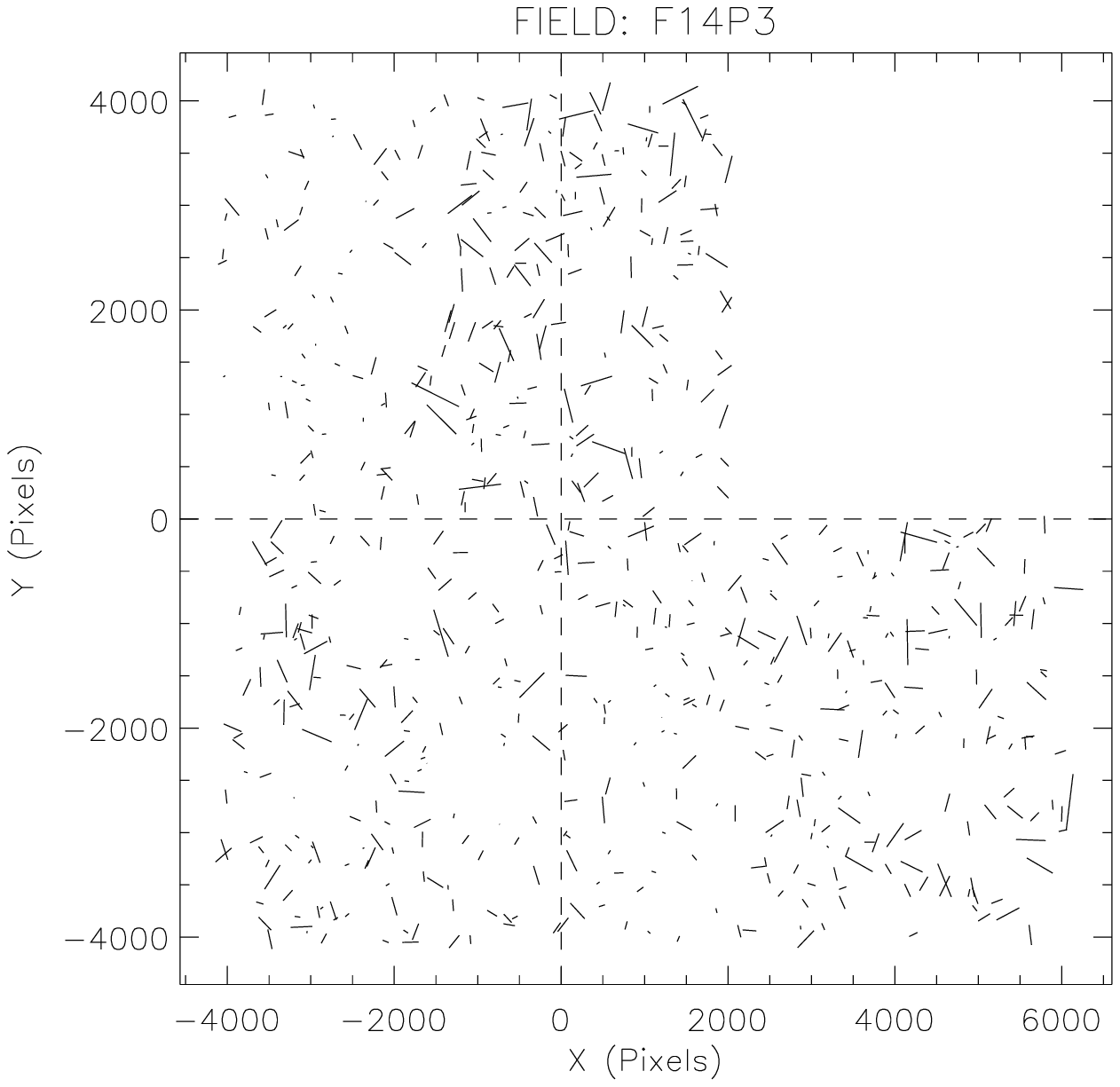,height=8cm}}
\caption{\label{F14P3.ps} Same as Figure \ref{F14P1.ps} for FIELD F14P3.}
\end{figure*}

\begin{figure*}
\centerline{
\psfig{figure=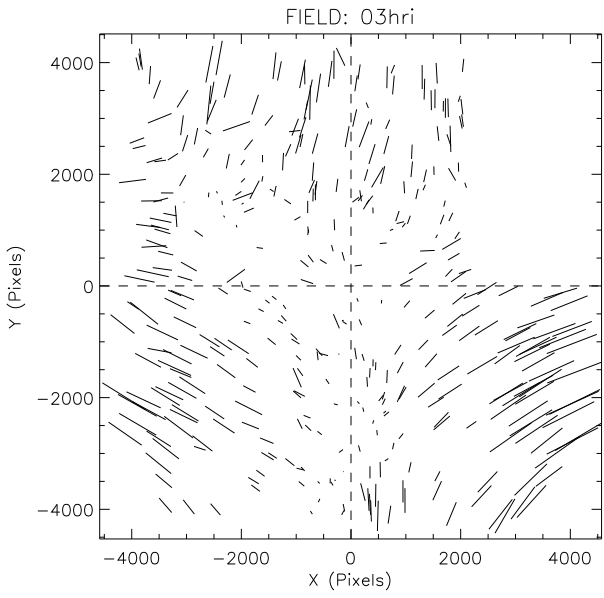,height=8cm}
\psfig{figure=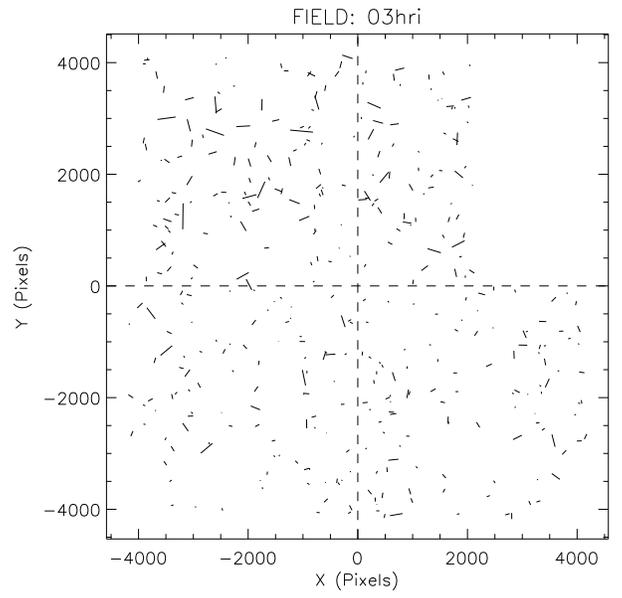,height=8cm}}
\caption{\label{03hri_e.ps} Same as Figure \ref{F14P1.ps} for FIELD 03hrie.}
\end{figure*}

\begin{figure*}
\centerline{
\psfig{figure=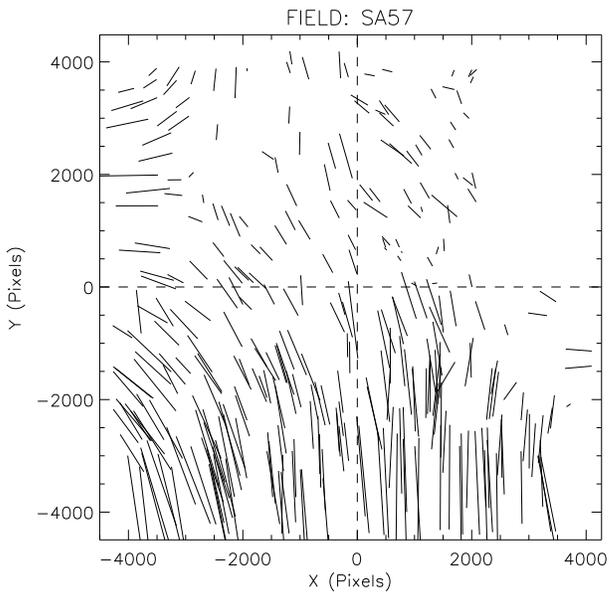,height=8cm}
\psfig{figure=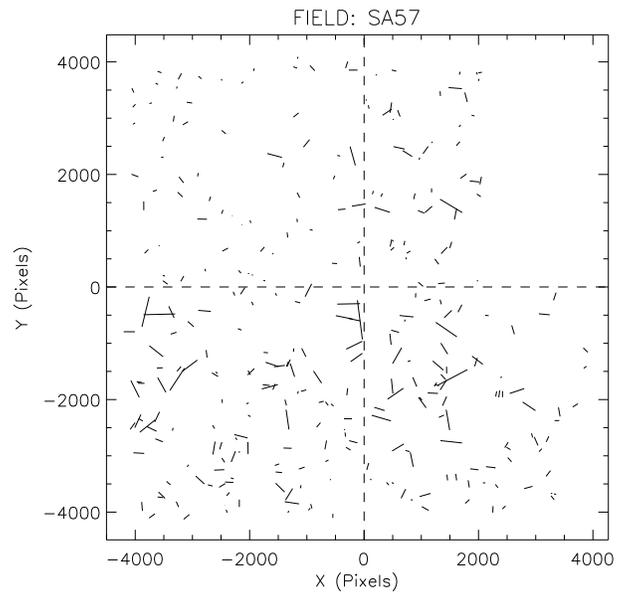,height=8cm}}
\caption{\label{SA57.ps} Same as Figure \ref{F14P1.ps} for FIELD SA57.}
\end{figure*}

\begin{figure*}
\centerline{
\psfig{figure=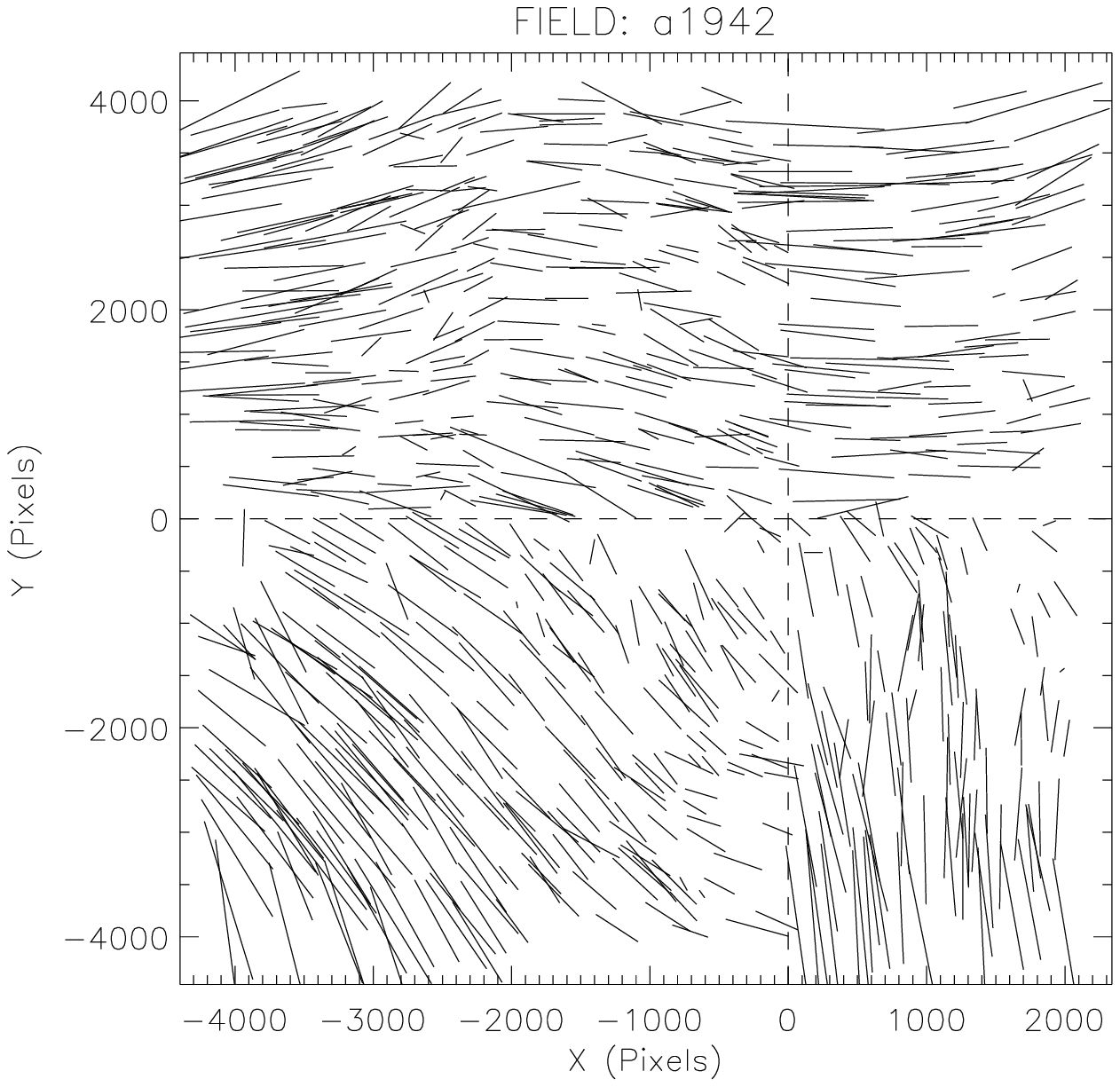,height=8cm}
\psfig{figure=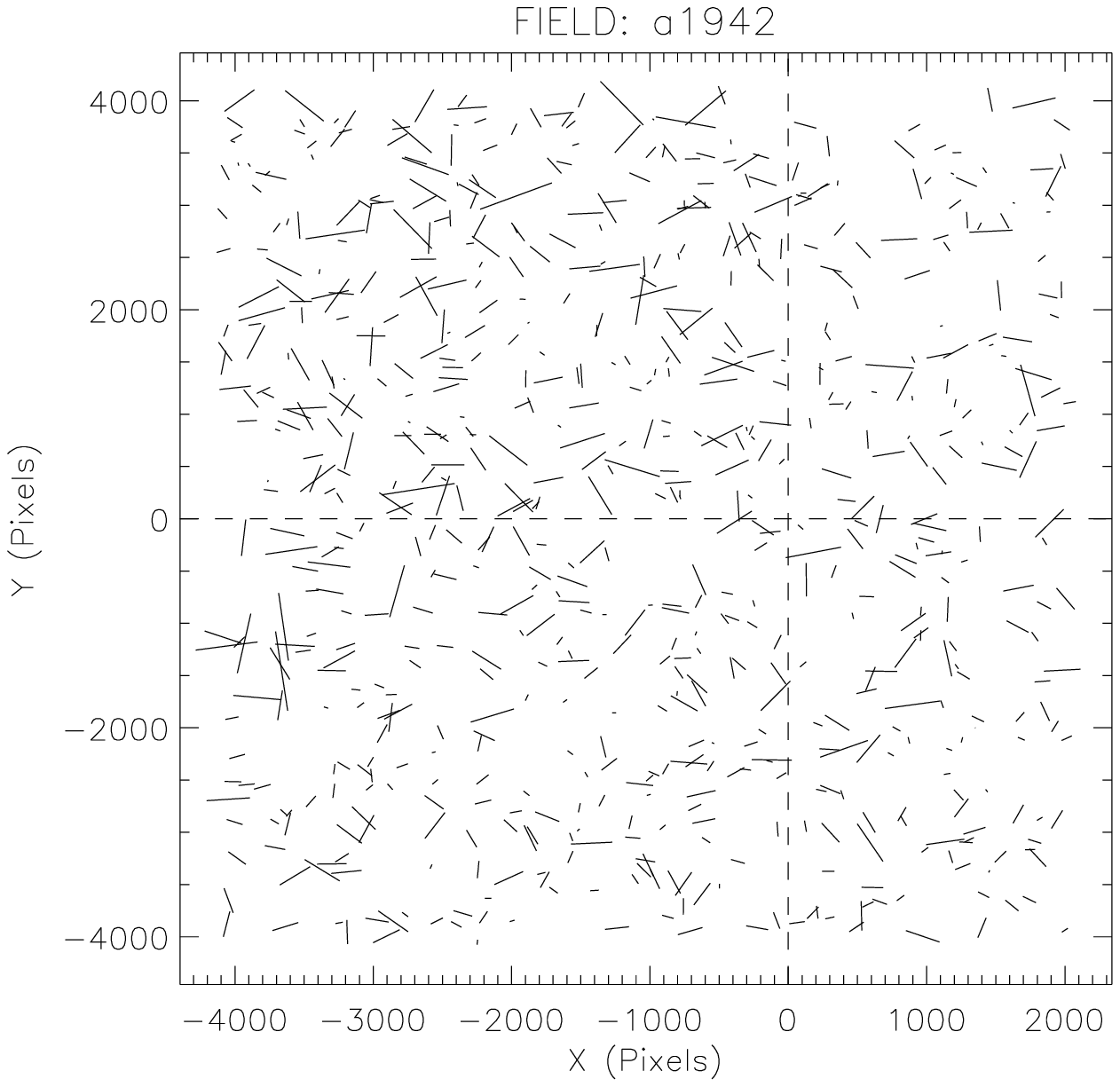,height=8cm}}
\caption{\label{a1942.ps} Same as Figure \ref{F14P1.ps} for FIELD a1942.}
\end{figure*}

\begin{figure*}
\centerline{
\psfig{figure=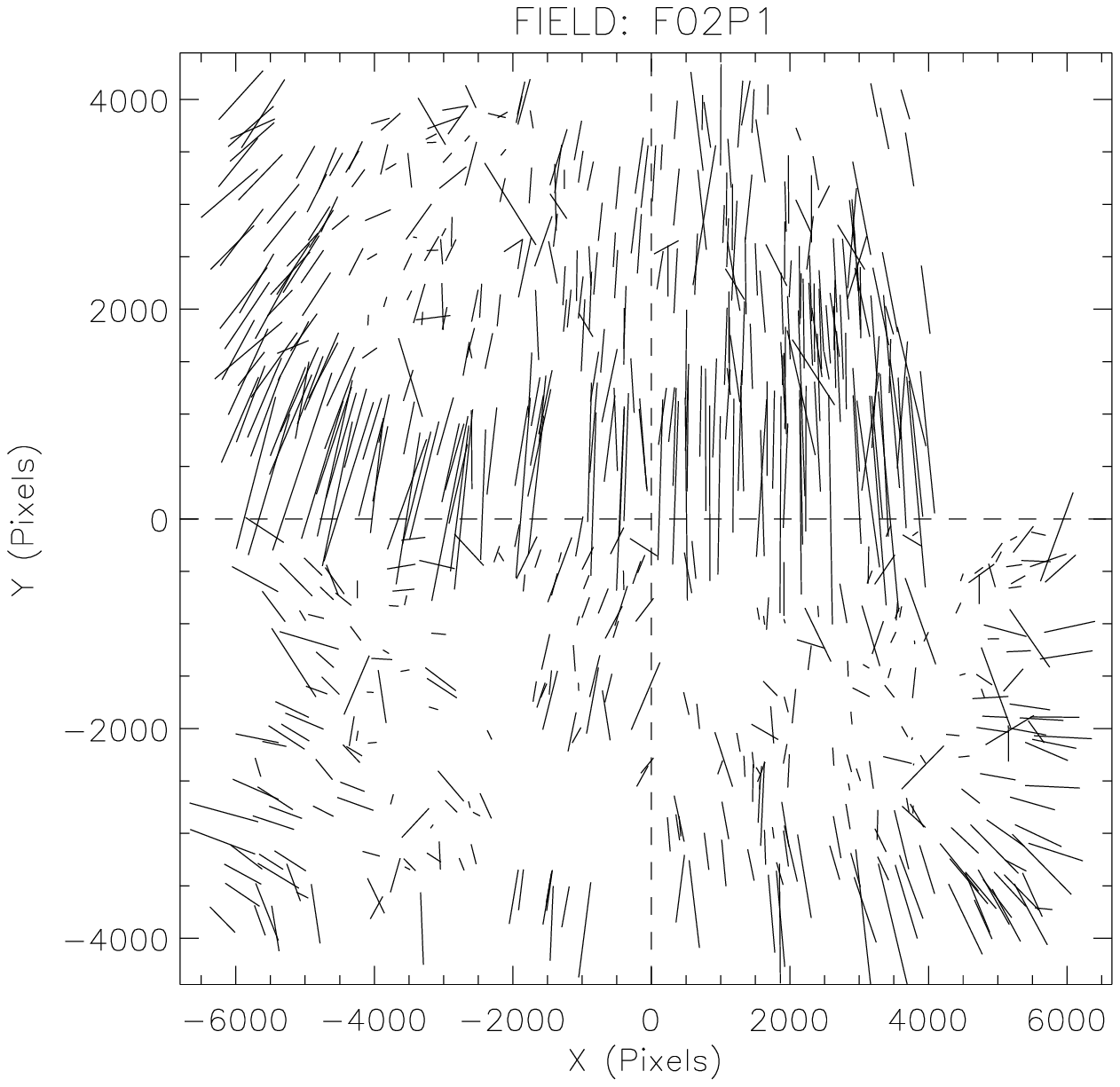,height=8cm}
\psfig{figure=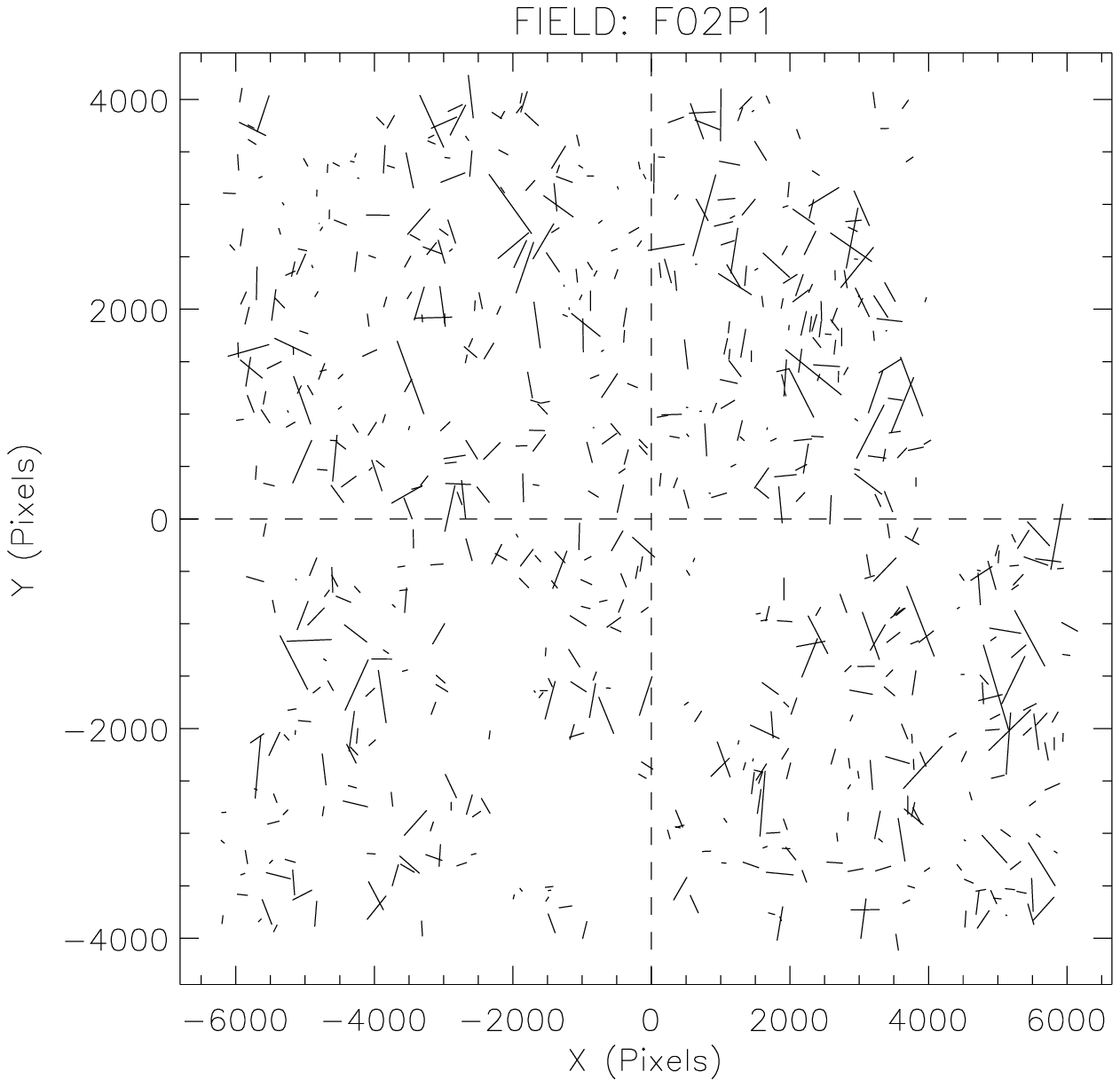,height=8cm}}
\caption{\label{F02P1.ps} Same as Figure \ref{F14P1.ps} for FIELD F02P1.}
\end{figure*}

\begin{figure*}
\centerline{
\psfig{figure=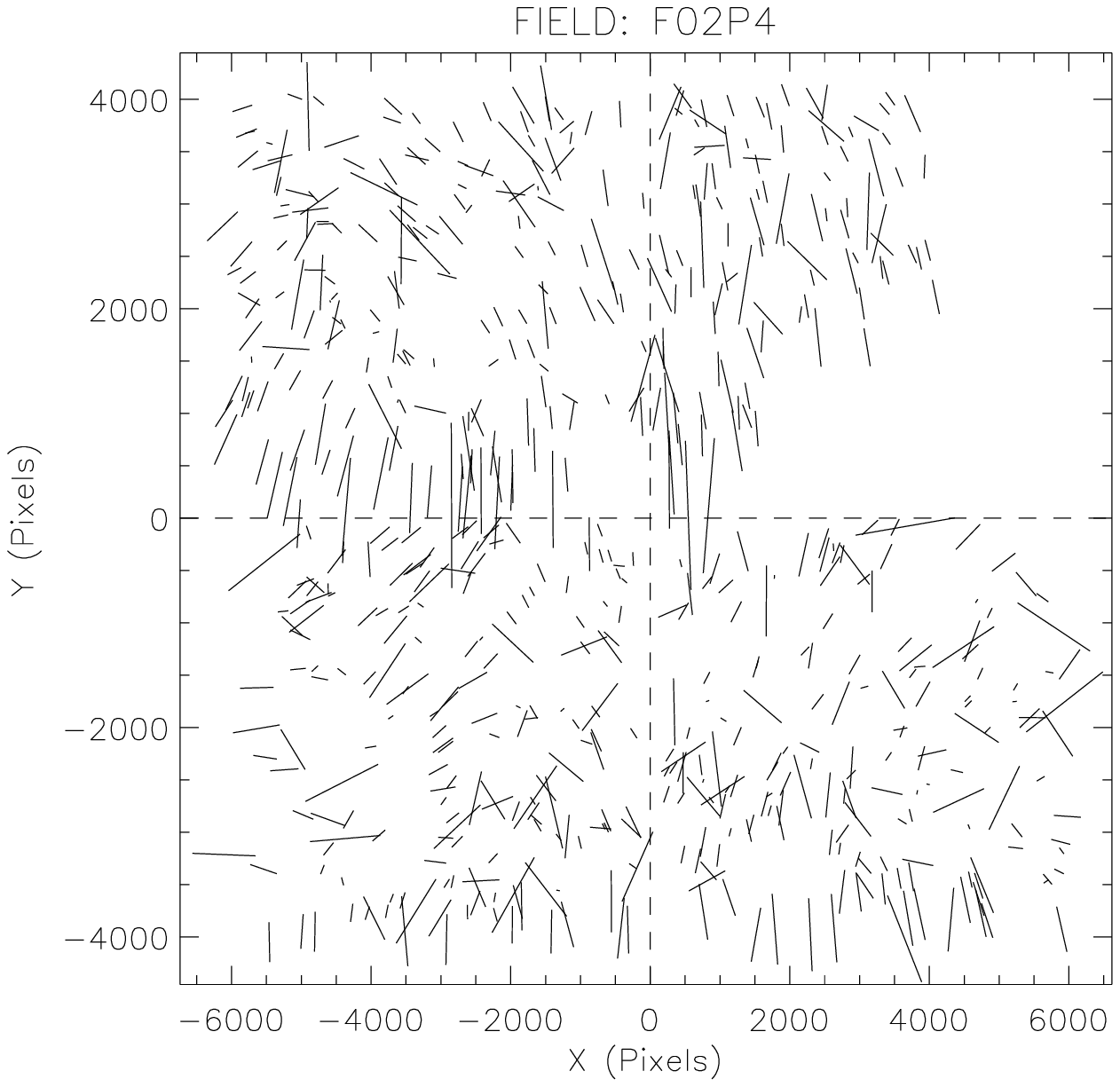,height=8cm}
\psfig{figure=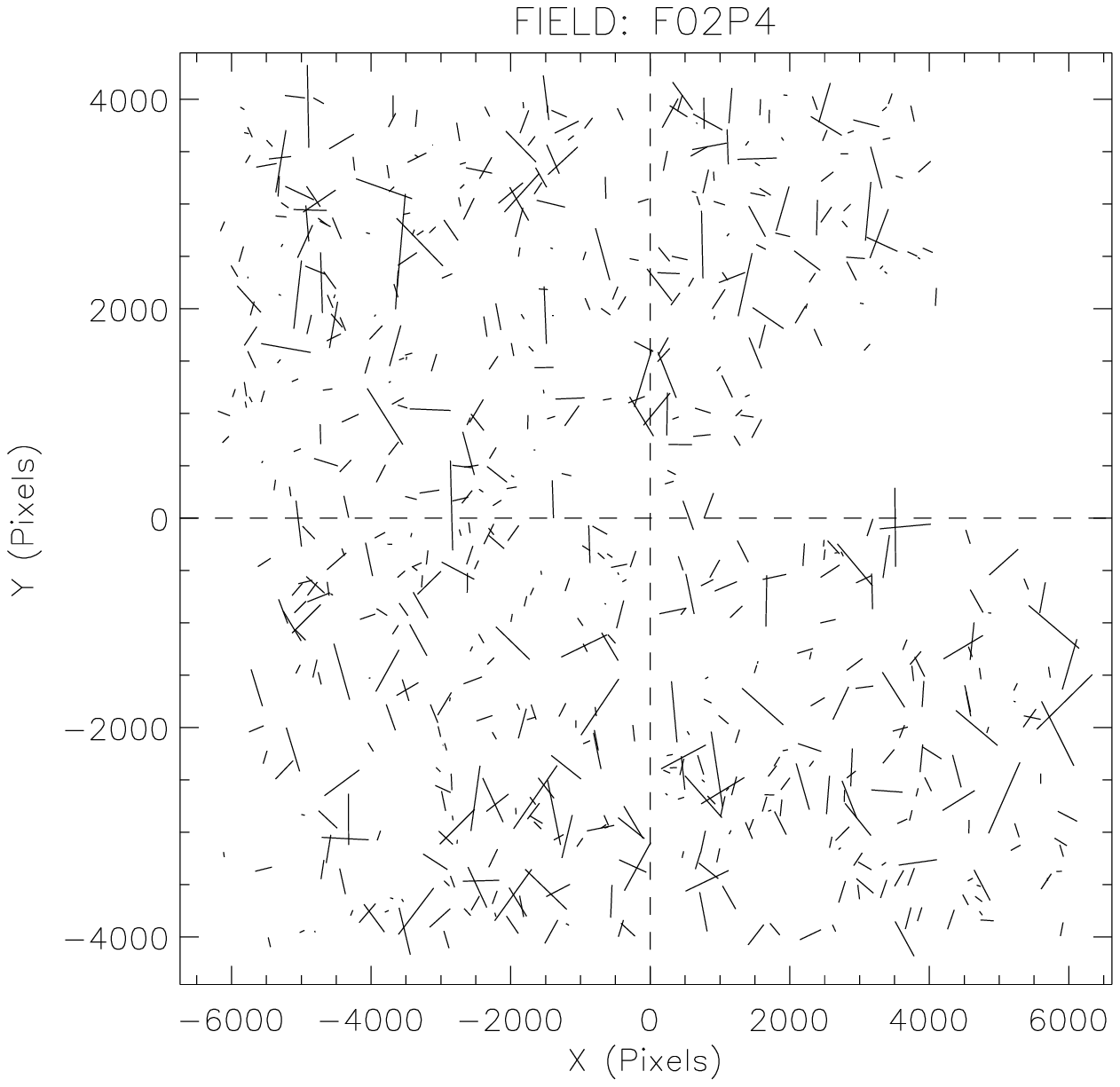,height=8cm}}
\caption{\label{F02P4.ps} Same as Figure \ref{F14P1.ps} for FIELD F02P4.}
\end{figure*}

We next study the second point  concerning
the residual of the PSF correction. Figure \ref{9700.f11.ps} shows
that the star ellipticity correction is efficient in removing PSF
anisotropies.
The raw star ellipticity can be as large as $20\%$ in the most extreme cases.
Figures \ref{F14P1.ps} to \ref{F02P4.ps} show maps of the uncorrected and the 
corrected star ellipticities. The same camera used at different times clearly
demonstrates that the PSF structure can vary a lot in both amplitude and
orientation, and that it is not dominated by the optical
distortion (as we can see from the location of the optical center, given by 
the
dashed cross). Individual CCD's
are $2K\times 4K$ chips, easily identified by the discontinuities in the stellar
ellipticity fields.

Next, let us sort the galaxies according to the increasing stellar ellipticity,
and bin this galaxy catalogue such that each bin contains a large number of galaxies.
We then measure, for each galaxy bin, two different
averaged galaxy ellipticities $\langle e_\alpha\rangle$: one is given by
Eq.(\ref{imcateq}) and the other by
Eq.(\ref{imcateq}), without the anisotropy correction $P^{sm} \pg$. The 
former should be uncorrelated with
the star ellipticity if the PSF correction is correct (let us call $\langle
e_\alpha\rangle$ this average); and
the latter should be strongly correlated with the star ellipticity, (let us
call $\langle e^{\rm ani}_\alpha\rangle$ this average).
Since the galaxies are binned according to the stellar
ellipticity, galaxies of a given bin are taken from everywhere in the 
survey, therefore the cosmic shear
signal should vanish, and the remaining possible non vanishing value for
$\langle e_1\rangle$ and $\langle e_2\rangle$
should be attributed to a residual of star anisotropy.
Figure \ref{9700.f4.ps} shows $\langle e_1\rangle$ and $\langle 
e_2\rangle$
(dashed lines) and
$\langle e^{\rm ani}_1\rangle$ and $\langle e^{\rm ani}_2\rangle$ (solid 
lines)
versus respectively
$\langle e^{\rm stars}_1\rangle$ and $\langle e^{\rm stars}_2\rangle$. The 
solid
lines exhibit a direct correlation between
the galaxy and the star ellipticities, showing that the PSF anisotropy
does indeed induce a strong spurious
anisotropy in the galaxy shapes of a few percents. However, the dashed lines
show that the corrected galaxy ellipticities
are no longer correlated with the star ellipticity, the average $\langle
e_1\rangle$ fluctuates around $-1\%$, while
$\langle e_2\rangle$ is consistent with zero. This figure shows the 
remarkable
accuracy of the PSF correction method
given in KSB. Error bars in these plots are calculated assuming Gaussian 
errors
for the galaxies in a given bin.
The significant offset of $\langle e_1\rangle$ of $1\%$ might be interpreted 
as a systematic induced by the CCD, as we
will see in the next Section, and can be easily corrected for.
Figure \ref{9700.f5.ps} shows that this systematic is nearly galaxy independent, and
affect all galaxies in the same way. This is also in favor of the CCD-induced systematic,
since we expect that a PSF-induced systematic (which is a convolution)
would depend on the galaxy size.

Figure \ref{9700.f6.ps} shows the same kind of analysis, but
instead of sorting the galaxies according to the star ellipticity amplitude,
galaxies are now sorted according to the distance $r$ from the optical 
center.
   The average quantities we measure are no longer
$\langle e_1\rangle$ and $\langle e_2\rangle$ versus $\langle e^{\rm
stars}_1\rangle$ and $\langle e^{\rm stars}_2\rangle$,
but the tangential and the radial ellipticity $\langle e_t\rangle$ and 
$\langle
e_r\rangle$ versus $r$. This new
average is powerful for extracting any systematic associated with the 
optical
distortion. Figure \ref{9700.f6.ps}
shows that the systematics caused by the optical distortion are a negligible 
part
of the anisotropy of the PSF, as
we should expect from Figures \ref{F14P1.ps} to \ref{F02P4.ps} (where the 
PSF
anisotropy clearly does not
follow the optical distortion pattern).

\subsection{Systematics due to the CCD frames}

\begin{figure}
\centerline{
\psfig{figure=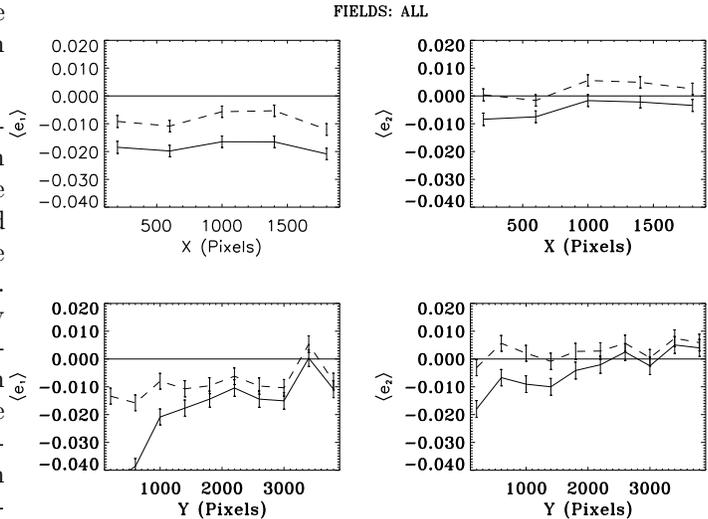,height=7cm}}
\caption{\label{9700.f7.ps} Average galaxy ellipticity $\langle
e_\alpha\rangle$
versus the $X$ and $Y$ location on the CCDs.
As for Figure \ref{9700.f4.ps}, the dashed lines are obtained from the 
fully
corrected galaxy ellipticities, as given by Eq.(\ref{imcateq}), and the 
solid
lines are obtained
from the corrected galaxy ellipticities where the anisotropy correction
term $P^{\rm sm}\pg$ has been removed from Eq.(\ref{imcateq}).
The systematic negative mean value of
$\langle e_1\rangle$ along lines or columns of the CCD (the two left panels)
show that the galaxies are preferentially aligned with the columns of the 
CCD
in the whole survey. A positive systematic value for $\langle e_2\rangle$ 
(the
two right panels) is also visible, although much less significant.
}
\end{figure}

Using the same method as in the previous Section, we can
also investigate the systematics associated with
the CCD line/columns orientations. Here, instead of sorting the galaxies
according to the star ellipticity or the
distance from the optical center, the galaxies are sorted according to their 
$X$
or $Y$ location on
each CCD frame. By averaging the galaxy ellipticities $\langle e_1\rangle$ 
and
$\langle e_2\rangle$
in either $X$ or $Y$ bins, we also suppress the cosmic shear signal and keep
only the systematics associated
with the CCD frame. Figure \ref{9700.f7.ps} shows $\langle e_1\rangle$ and
$\langle e_2\rangle$ (dashed lines)
and $\langle e_1\rangle$ and $\langle e_2\rangle$ (solid lines) versus 
$\langle
X\rangle$ and $\langle Y\rangle$.
The plots from the top-left to bottom-right correspond respectively to 
$\langle
e_1
\rangle$ versus $\langle X\rangle$, $\langle e_2 \rangle$ versus $\langle
X\rangle$, $\langle e_1\rangle$ versus
$\langle Y\rangle$, and $\langle e_2 \rangle$ versus
$\langle Y\rangle$. We see that
$\langle e_1\rangle$ is systematically negative by $\sim -1\%$ for
both X and Y binnings,
while $\langle e_2 \rangle$ does not show any significant deviation from 
zero.
This result is fully consistent
with the dashed lines in Figure \ref{9700.f4.ps} which demonstrate that 
the
$-1\%$ systematic is probably
a constant systematic which affects all the galaxies in the same way, and 
which
is not related to the
star anisotropy correction. The origin of this constant shift is still not
clear, it might have been produced
during the readout process, since a negative $\langle e_1\rangle$ 
corresponds
to an anisotropy along columns of the CCDs.

\subsection{Test of the systematics residuals}

\begin{figure}
\centerline{
\psfig{figure=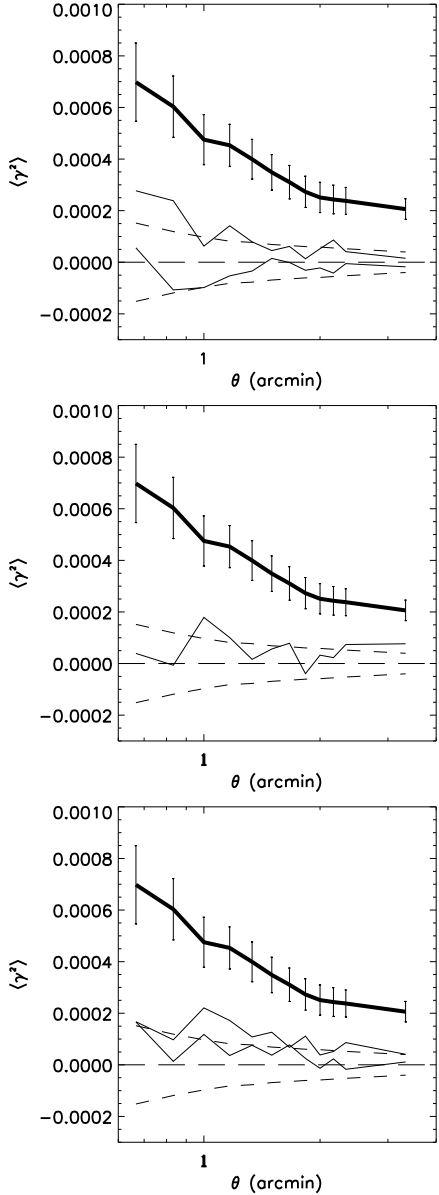,height=16cm}}
\caption{\label{9700.f8.ps} Possible contribution of the systematics
studied in Section 3.2 to the signal. On each of the plots, the thick solid 
line
shows the signal as displayed on Figure \ref{9700.f1.ps}, and the
dashed lines show the $\pm 1 \sigma$ fluctuation obtained from $1000$
random
realizations. From top to bottom:
(a) The two thin solid lines are $\langle\gamma^2\rangle^{1/2}$ measured
on the galaxies sorted according to the star ellipticity strength (see 
Figure
\ref{9700.f4.ps}). For the different smoothing scales, the mean number
and the variance of the number of galaxies in the chosen bins fit the one
observed in the signal (thick solid) curve. (b) the thin solid line is
$\langle\gamma_t^2\rangle^{1/2}$ measured on the galaxies sorted according
to the distance from the optical center, and on (c) the two thin solid lines
correspond to the galaxies sorted
according to their $X$ or $Y$ location on the CCDs.}
\end{figure}

The correction of the constant shift of $-1\%$ along $\langle e_1\rangle$ 
has been applied to the galaxy catalogue
from the beginning. It ensures that there is no more significant residual
systematic (either star anisotropy
or optical distortion or CCD frame), and demonstrates that the average level
of residual systematics is small and much
below the signal. However we have to check that the systematics do not oscillate
strongly around this small value. If it were the case, then this small level of
systematics could still contribute significantly to the variance of the shear. This
can be tested by calculating the variance of the shear in bins much smaller than those
used in Figure \ref{9700.f4.ps} to calculate the average level of residual systematics.
In order to decide how small the bins should be
we can use the number of galaxies available in the apertures used to measure the signal,
for a given smoothing scale. For example for $\theta=0.66'$ there is 45 galaxies in average,
and for $\theta=3.3'$ there is 1100 galaxies. We can therefore translate a bin size into
a smoothing scale, via the mean number of galaxies in the aperture.
We found that the variance of the shear measured in these smaller bins is still
negligible with respect to the signal, as shown by Figure \ref{9700.f8.ps}.
The three panels from top to bottom show respectively the star anisotropy 
case, the optical distortion case and the CCD frame case. On each panel,
the thick solid line is the signal with its error bars
derived from $1000$ randomizations. The short dashed lines show the $\pm
1\sigma$ of these error bars
centered on zero. On the top panel the two thin solid lines show $\langle
\gamma^2\rangle$ respectively
measured with the galaxies sorted according to $e^{\rm stars}_1$ and to 
$e^{\rm
stars}_2$. The thin solid line
in the middle panel shows $\langle \gamma_t^2\rangle$ measured from the 
galaxies,
sorted according to their distance from
the optical center, and the two thin solid lines in the bottom panel show
$\langle \gamma^2\rangle$ measured on
the galaxies sorted according to $X$ and $Y$.

In all the cases, the thin solid lines are consistent with the $\pm 1\sigma$
fluctuation, without showing
a significant tendency for a positive $\langle \gamma^2\rangle$. We conclude
that the residual systematics
are unable to explain the measured $\langle \gamma^2\rangle$ in our survey, 
and
therefore our signal is likely to be of cosmological origin.

\begin{figure}
\centerline{
\psfig{figure=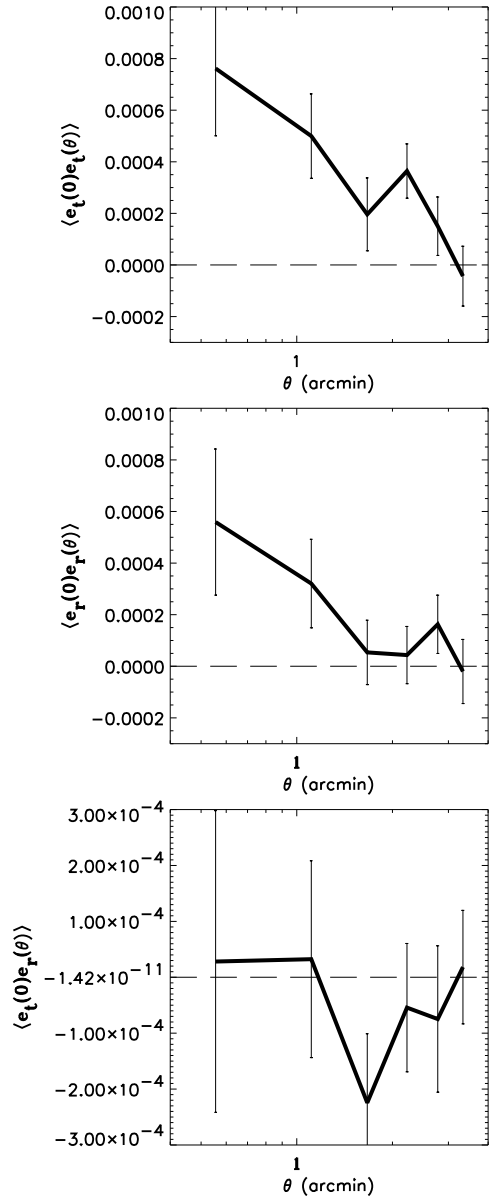,height=16cm}}
\caption{\label{9700.f9.ps} From top to bottom, measurement of the
correlation functions $\langle e_t(0) e_t(\theta)\rangle$, $\langle e_r(0)
e_r(\theta)\rangle$
and $\langle e_r(0) e_t(\theta)\rangle$. The error bars are computed from 50
random realizations
of our data set where the orientations of the galaxies were randomized.}
\end{figure}

A direct test of its cosmological origin is to measure the correlation 
functions
$\langle
e_t(0) e_t(\theta)\rangle$, $\langle e_r(0) e_r(\theta)\rangle$ and $\langle
e_r(0)
e_t(\theta)\rangle$, where $e_t$ and $e_r$ are the tangential and radial
component
of the shear respectively:

\begin{eqnarray}
e_t=-e_1~\cos(2\theta_k)-e_2\sin(2\theta_k)\nonumber\\
e_r=-e_2~\cos(2\theta_k)+e_1\sin(2\theta_k),
\end{eqnarray}
where $\theta_k$ is the position angle of a galaxy. If the signal is
due to gravitational shear, we can
show (\cite{K92}) that $\langle e_t(0) e_t(\theta)\rangle$ should be 
positive,
$\langle e_r(0)
e_r(\theta)\rangle$ should show a sign inversion at intermediate scales, and
$\langle e_r(0)
e_t(\theta)\rangle$ should be zero. This is a consequence of the scalar 
origin
of the gravitational
lensing effect and of the fact that galaxy ellipticity components are
uncorrelated. Although we do not yet
have enough data to perform an accurate measurement of these
correlation functions,
it is interesting to check their general behavior.
Figure \ref{9700.f9.ps} shows that
in our data set, although the measurement is very noisy, both $\langle 
e_t(0)
e_t(\theta)\rangle$
and $\langle e_r(0) e_r(\theta)\rangle$ are positive valued, while $\langle
e_r(0) e_t(\theta)\rangle$
is consistent with zero. This measurement demonstrates that the component of 
the
galaxy ellipticities
$e_\alpha$ of well separated
galaxies are uncorrelated, and it is in some sense a strong indication that 
our
signal at small
scales is of cosmological origin.

The last thing we have checked is the
stability of the results with respect to the field selection.
  We verified that removing one of the fields consecutively for all the
fields
(see Section 2 for the list of the fields) does not
change the amplitude and the shape of the signal, even
for the Abell 1942 field. The cluster has
  no impact and does not bias the analysis because it was significantly
  offset from the optical axis. This  ensures that
the signal is not produced by one field
only, and that they are all equivalents in terms of image quality,
PSF correction accuracy and signal amplitude, even using V and I colors. It
also validates the different
pre-reduction methods used for the different fields.

\section{Cosmological constraints}

Figure \ref{9700.f1.ps} provides a first comparison of our signal with
some cosmological models. In order to rule out  models we need to
  estimate first the
sample variance in the variance of the shear. Although it
has not been yet exactly derived
analytically  (because calculations in the non-linear regime are difficult),
ray-tracing simulations can give an accurate estimate of it. We used the 
ray-tracing
simulations
of \cite{JSW99} for this purpose.

\begin{figure}
\centerline{
\psfig{figure=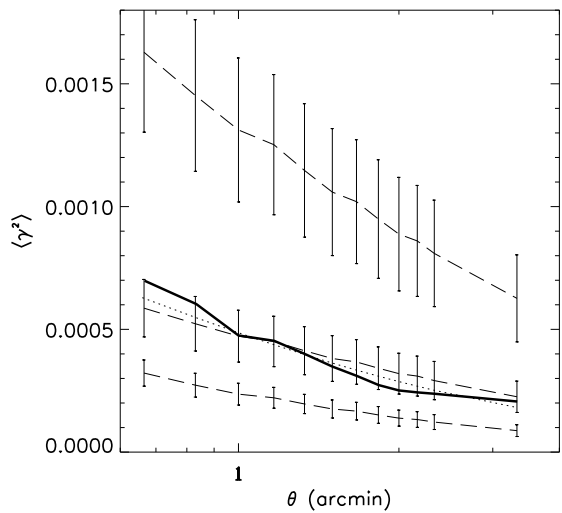,height=7cm}}
\caption{\label{9700.f10.ps} Comparison of our signal (thick line) 
with
three cosmological
models. The error bars are the cosmic variance measured on five independent
realizations
at the smoothing scale indicated by the x-axis. For clarity, the shot noise
error bars of the
signal are not
plotted, their amplitude can be read on Figure \ref{9700.f8.ps}. From
bottom to top, the dashed lines correspond to: model (1), model (2) and 
model
(3) as given in Table
\ref{simul}. The shot-noise error bars of the signal are in fact comparable 
in
amplitude
to the cosmic variance error bars of model (2). We show also a
cluster-normalized
$\Lambda$ model (dotted line) with $\Omega=0.3$, $\Lambda=0.7$, and a CDM
power spectrum with $\Gamma=0.5$. This model was not obtained from a
simulation, but computed using the non-linear power spectrum using the
\cite{PD96} formula.}
\end{figure}

\begin{table}
\caption{List of the ray tracing simulations we used (see \cite{JSW99} for
details).
The redshift of the sources is $1$.
}
\label{simul}
\begin{center}
\begin{tabular}{|c|c|c|c|c|c|}
\hline
Simulation \# & $\Gamma$ & $\Omega_0$ & $\Lambda$ & $\sigma_8$ \\
\hline
(1) OCDM       & 0.21 & 0.3 & 0 & 0.85 \\
(2) $\tau$CDM  & 0.21 & 1   & 0 & 0.6 \\
(3) $\tau$CDM  & 0.21 & 1   & 0 & 1 \\
\hline
\end{tabular}
\end{center}
\end{table}
Table \ref{simul} shows the two simulations we used. The $\tau$CDM model 
with
$\sigma_8=1$ is
not an independent simulation, but was constructed from the $\tau$CDM model 
with
$\sigma_8=0.6$ simply by dividing $\kappa$ by $0.6$.
This should empirically mimic a model with both $\Omega_0$ and $\sigma_8$
equal to one. The redshift of the sources is equal to $1$, which is
not appropriate
for our data. However, for the depth of the survey, we believe that it
represents fairly well the mean redshift of the galaxies, which is the 
dominant
factor in determining the second moment.
Figure \ref{9700.f10.ps} shows the amplitude and the scale 
dependence
of the variance of the
shear for the three cosmological models, compared to our signal. It is
remarkable that models
(1) and (3) can be marginally rejected
(We did not plot the error bars due to the intrinsic ellipticity for
clarity: they can be obtained from Figure \ref{9700.f8.ps}).

  Our measurements are in agreement with the cluster normalized model (2).
Also plotted is the theoretical prediction of a $\Lambda$CDM model, with
$\Omega=0.3$, $\Lambda=0.7$, $\Gamma=0.5$ and a redshift of the sources
$z_s=1$. It shows that the low-$\Omega$ model is also in good agreement
with the data, which means that weak gravitational lensing provides
cosmological constraints similar to the cluster abundance results
(\cite{E96},\cite{B99}):
the second moment of the shear measures a combination of $\sigma_8$
and $\Omega_0$ (see equation \ref{vartheo}). A measure of the
third moment of the convergence would break the $\Omega$-$\sigma_8$
degeneracy, but this requires more data (see \cite{B97}, \cite{VW99}, 
\cite{JSW99}).
It should also be noted that for the simulations,
we have considered cold dark matter
models with shape parameter $\Gamma=0.21$; higher values of $\Gamma$
increase the theoretical predictions on scales of interest, e.g. the 
$\Omega_0=1$,
$\sigma_8=1$ model would be ruled out even more strongly.
  We conclude
that our analysis is consistent with the current favored cosmological
models, although we
cannot yet  reject other models with high significance.
Since we have only analyzed 2 square degrees of the survey,
with forthcoming larger surveys we should be able to
set strong constraints on the cosmological models as discussed below.

Due to the imprecise knowledge of the redshift
distribution in our data, the interpretation might still be subject to
modifications. The final state of our survey in 4 colors will however
permit the measurement of this distribution by estimating photometric
redshifts for the source galaxies.

\section{Conclusion}

We have demonstrated the existence of a significant correlation between 
galaxy
ellipticities
from 0.5 to 3.5 arc-minutes scales. The signal
has the amplitude and the angular dependence
expected from theoretical predictions of
weak lensing produced by large-scale structures in the universe.
  We have tested the possible contribution of systematic errors
to the measured signal; in
particular we discussed three potential sources of spurious alignment of
galaxies: overlapping isophotes
of very close galaxies, star anisotropy and CCD line/column alignment. The 
first
of these systematics
is easy to deal with, simply by removing close pairs,
although we
  may have decreased the signal slightly by removing them.
The star anisotropy seems
to be very well controlled, in part due to the fact that the
bias adds quadratically with the
signal. Moreover, in the absolute sense, the bias does not exceed
a fraction of 1 percent, which is
adequate to accurately measure a variance of the shear of few
percent.
The only important bias we found seems to be associated with the CCD 
columns,
and it is constant over
the survey, it is therefore easy to correct for. The origin of this CCD bias 
is
still unclear.

As an objective test of the reality of the gravitational shear signal,
we measured the ellipticity correlation functions
$\langle e_t(0) e_t(\theta)\rangle$,
$\langle e_r(0) e_r(\theta)\rangle$ and
$\langle e_r(0) e_t(\theta)\rangle$. While the measurement is
noisy, the general behavior is fully consistent with the lensing
origin of the signal.
The tests for systematic errors and the three ellipticity
correlation function measurements described above have led us to
conclude with confidence that we have measured a cosmic shear signal.

With larger survey area, we expect
to be able to measure other lensing statistics, like
the aperture mass statistic ($M_{\rm ap}$;
see \cite{SVWJK98}). The $M_{\rm ap}$ statistic is still
very noisy for our survey size because
its signal-to-noise is lower than the top-hat smoothing
statistic, due to higher sample variance
(We verified this statement using the ray tracing simulation data
of \cite{JSW99}). Our
survey will increase in size in the near future
(quickly up to $7$ square degrees), leading to a factor of 2 improvement in
the signal-to-noise of the results presented here. According to our 
estimates,
this will be enough to measure $M_{\rm ap}$ at the arc-minute scale with
a signal-to-noise of $\sim 3$.
The detection of the skewness of the convergence
should also be possible with the increased survey area. This will
be important in breaking the degeneracy between the amplitude of
the power spectrum and $\Omega$ (\cite{B97}, \cite{VW99}, \cite{JSW99}).
These measures should also provide nearly independent
confirmations of the weak gravitational lensing effect as well
as additional constraints on cosmology.
Thus by combining different measures
of lensing by large-scale structure
(top-hat smoothing statistics, $M_{\rm ap}$ statistics, correlation
function analysis, power spectrum measurements), higher order
moments, and peak statistics (\cite{JVW00}),  from
forthcoming survey data, we hope to make significant progress in
measuring dark matter clustering and cosmological parameters with
weak lensing.

We also hope to do a detailed analysis
with a more sophisticated PSF correction algorithm. For instance, the mass
  reconstruction is linear with the amplitude of the residual
  bias, and a fraction of percent bias is still
  enough to prevent a definitive detection of filaments
or to map the details of large scale structures. Since
  we show elsewhere (\cite{E00}) that such a bias is unavoidable with
the present day correction techniques and image quality, there is still
room to improve the analysis prior to obtaining accurate large-scale mass 
maps.
Recent efforts to improve the PSF correction are very encouraging 
(\cite{Ks99}).
  We plan to explore such approaches once we get
   an essentially homogeneous data set on a larger field.

{
\acknowledgements T. Erben and R. Maoli thank CITA for hospitality,
L. Van Waerbeke thanks IAP and MPA for hospitality,
F. Bernardeau, J.-C. Cuillandre and T. Erben
  thank IAP for hospitality.  We thank
  M. Cr\'ez\'e and A. Robin for providing their UH8K data of SA57.
We thank P. Couturier for the allocation of CFHT discretionary
time to observe the Abell 1942 cluster with UH8K.
  This work was supported by the TMR Network ``Gravitational Lensing: New
Constraints on
Cosmology and the Distribution of Dark Matter'' of the EC under contract
No. ERBFMRX-CT97-0172, and a PROCOPE grant No. 9723878 by the DAAD and
the A.P.A.P.E. We thank the TERAPIX data center for providing its facilities
for the data reduction of the CFH12K and UH8K data.  
}

\end{document}